\documentclass[sigconf]{acmart}

\AtBeginDocument{%
  }

\copyrightyear{2022} 
\acmYear{2022} 
\setcopyright{acmcopyright}\acmConference[APCCPA '22]{Proceedings of the 1st International Workshop on Advances in Point Cloud Compression, Processing and Analysis}{October 14, 2022}{Lisboa, Portugal}
\acmBooktitle{Proceedings of the 1st International Workshop on Advances in Point Cloud Compression, Processing and Analysis (APCCPA '22), October 14, 2022, Lisboa, Portugal}
\acmPrice{15.00}
\acmDOI{10.1145/3552457.3555727}
\acmISBN{978-1-4503-9491-8/22/10}

\usepackage{balance}
\usepackage{multirow}
\usepackage{subfigure}

\def\ie{{\it i.e.}}
\def\eg{{\it e.g.}}
\def\etc{{\it etc.}}
\def\etal{{\it et~al.}}

\begin{document}

\title[GRASP-Net]{GRASP-Net: Geometric Residual Analysis and Synthesis for Point~Cloud~Compression}


\author{Jiahao Pang}
\affiliation{%
  \institution{InterDigital}
  \city{New York City}
  \state{NY}
  \country{USA}}
\email{jiahao.pang@interdigital.com}

\author{Muhammad Asad Lodhi}
\affiliation{%
  \institution{InterDigital}
  \city{New York City}
  \state{NY}
  \country{USA}}
\email{muhammad.lodhi@interdigital.com}

\author{Dong Tian}
\affiliation{%
  \institution{InterDigital}
  \city{New York City}
  \state{NY}
  \country{USA}}
\email{dong.tian@interdigital.com}

\renewcommand{\shortauthors}{Jiahao Pang, Muhammad Asad Lodhi, \& Dong Tian}

\begin{abstract}
Point cloud compression (PCC) is a key enabler for various 3-D applications, owing to the universality of the point cloud format.
Ideally, 3D point clouds endeavor to depict object/scene surfaces that are continuous.
Practically, as a set of discrete samples, point clouds are locally disconnected and sparsely distributed.
This sparse nature is hindering the discovery of local correlation among points for compression.
Motivated by an analysis with fractal dimension, we propose a heterogeneous approach with deep learning for lossy point cloud geometry compression.
On top of a base layer compressing a coarse representation of the input, an enhancement layer is designed to cope with the challenging geometric residual/details.
Specifically, a point-based network is applied to convert the erratic local details to latent features residing on the coarse point cloud.
Then a sparse convolutional neural network operating on the coarse point cloud is launched.
It utilizes the continuity/smoothness of the coarse geometry to compress the latent features as an enhancement bit-stream that greatly benefits the reconstruction quality.
When this bit-stream is unavailable, \eg, due to packet loss, we support a skip mode with the same architecture which generates geometric details from the coarse point cloud directly.
Experimentation on both dense and sparse point clouds demonstrate the state-of-the-art compression performance achieved by our proposal. 
Our code is available at https://github.com/InterDigitalInc/GRASP-Net.
\end{abstract}




\maketitle

\section{Introduction}
\label{sec:intro}
As the raw output of many affordable depth-sensing devices, point cloud has become an indispensable way to represent 3D geometry, with a wide spectrum of applications ranging from AR/VR, autonomous driving, to topography, \etc\ 
However, 3D point clouds in the real world contain a huge number of points (\eg, millions).
To alleviate the cost of transmission/storage, point cloud compression (PCC) becomes a critical phase for 3D applications.

\begin{figure}[t]
  \centering\scriptsize
  \subfigure[12-bit dense point clouds]{\includegraphics[height=107pt]{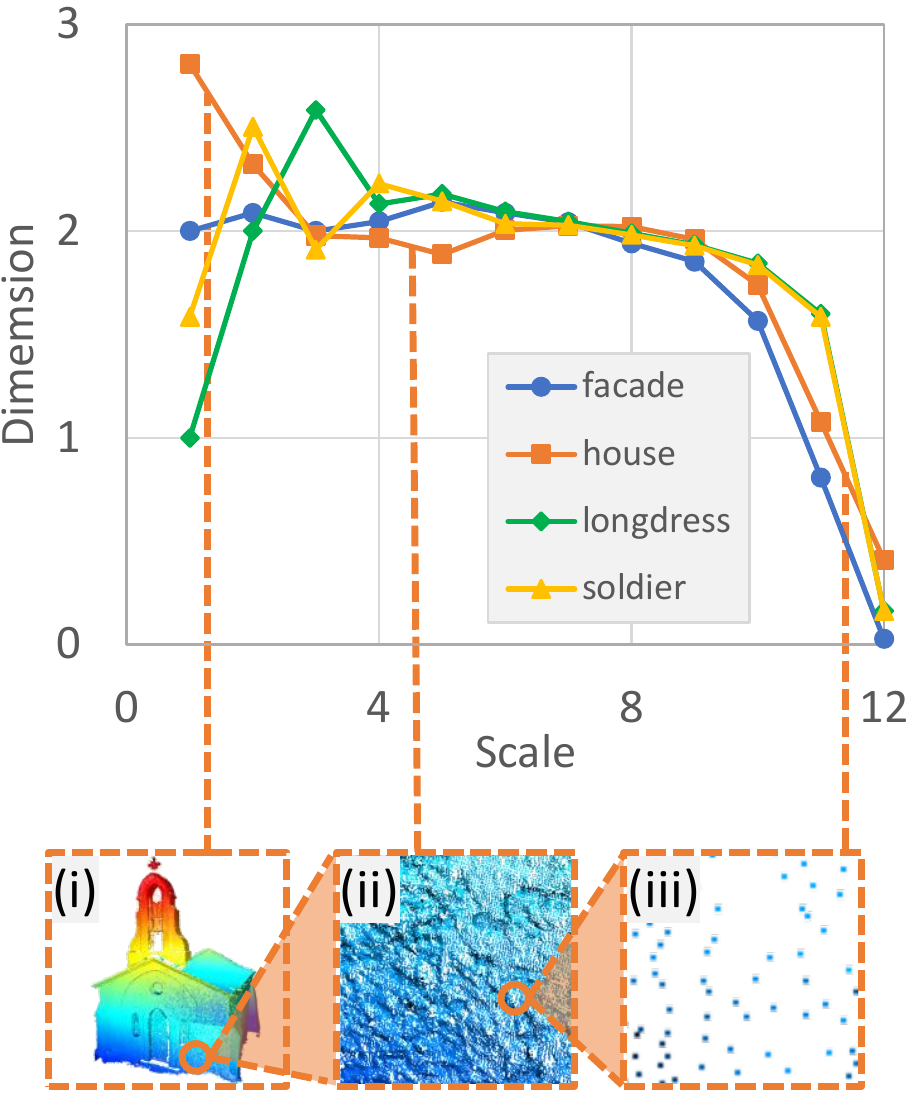}\label{fig:dim_dense}}\hspace{1pt}
  \subfigure[18-bit sparse LiDAR point clouds]{\includegraphics[height=107pt]{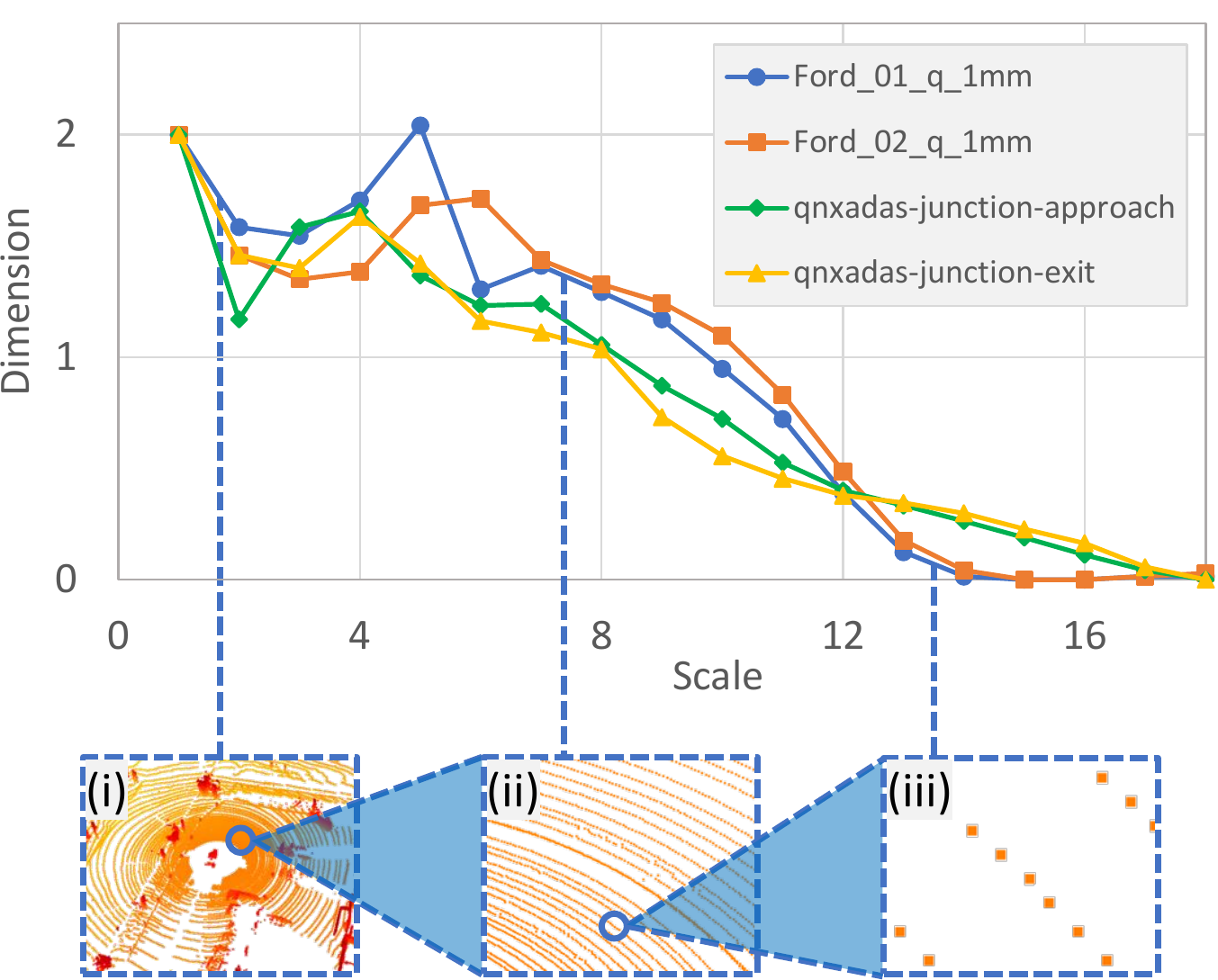}\label{fig:dim_sparse}}
  \vspace{-8pt}
  \caption{\small The ``effective'' dimensions of point clouds at different scales. Dimensions of dense point clouds (a) are $2$ at most scales as they represent 2-D surfaces, \eg, (a-ii); dimensions of LiDAR point clouds (b) are around $1$ due to the line patterns introduced by spinning LiDARs, \eg, (b-ii). However, they both degrade to $0$ at the finest scales, indicating local sparsity, \eg, (a-iii) and (b-iii). This observation motivates us to propose a heterogeneous PCC framework.}
  \label{fig:dim}
  \vspace{-14pt}
\end{figure}

A point cloud delineates the object (or scene) geometry by a set of 3D points.
Object and scene surfaces are \emph{continuous} manifolds embedded in the 3D space, such continuity/smoothness is favorable for compression (and other processing tasks) since it eases the discovery of local signal correlation.
However, as \emph{discrete} samples of surfaces, real-world point clouds are locally sparse and unordered.
Our analysis in Section~\ref{ssec:sparsity_analysis} reveals that the ``effective'' dimension of a point cloud gradually degrades to $0$ as bit-depth/scale increases (Fig.~\ref{fig:dim}).
In contrast to 2D natural images (or videos) where pixel patches reside in a dense Euclidean domain, the sparse nature of point clouds makes it highly challenging to capture the correlation among neighboring points.
Therefore, how could a PCC algorithm be designed to account for the local sparsity?
To shed light on this question, we revisit popular point cloud representations below.

\emph{Voxel} representation organizes 3D points as uniform voxel grids by quantizing the 3D coordinates, making them able to be processed like 2D images, \eg, with a convolutional neural network (CNN)~\cite{ioannidou2017deep}. 
However, a small quantization step size introduces many empty voxels due to local sparsity, which harms the propagation of information.
For instance, a convolution kernel may only touch one occupied voxel and make the processing highly inefficient.
Hence, a larger quantization step size is preferred; however, it comes at the cost of compromising the representability of intricate local details.
A related representation is the tree-based representation (such as an octree) which encapsulates the voxels as leaf nodes under a hierarchical tree structure \cite{schnabel2006octree}.
However, building a tree on top of the voxels does not resolve the difficulty in representing local geometric details.

On the contrary, the \emph{point-based} representation does not apply any pre-processing to the point clouds and preserves local geometric details.
Under the point-based representation, each point is directly specified by its $x$, $y$, and $z$ coordinates.
Though it is traditionally challenging to consume native 3D point sets as they are unordered, it is enabled with the recent development of point-based deep neural networks \cite{qi2017pointnet}.

The above observations motivate us to propose a \emph{heterogeneous} PCC solution.
We decompose a point cloud as a coarse component sketching the rough geometry, and the sparse local residual attached on top of it.
The coarse component has stronger continuity, making it suitable to be processed under the voxel/octree representations; while we rely on the point-based representation to tackle the sparsity of the local geometric residual.
The coarse component, represented as a voxelized point cloud, is first compressed with our base layer using tree-based coding.
On top of the base layer, an \emph{enhancement} layer is dedicated to compressing the geometric residual.
It not only tackles the local sparsity but also further exploits the continuity of the coarse point cloud for efficient compression.
The geometric residual is first represented as local point sets and fed to a point-based analysis network for extracting pointwise features of the coarse point cloud.
Then a feature codec based on sparse convolutions is launched for feature analysis and synthesis, leading to an enhancement bit-stream.
It operates on the voxelized coarse point cloud to further discover the spatial correlation among neighboring features.
In the end, the geometric residual is recovered with a point-based synthesis network and then added back to the coarse point cloud to complete the decoding.

When the enhancement bit-stream is unavailable, \eg, due to packet loss or network congestion, our decoder switches to a \emph{skip} mode which endeavors to generate the geometric residual solely from the coarse point cloud.
In this scenario, it behaves as a post-refinement to the coarse point cloud.
With our design, the network architecture remains unchanged under the skip mode.
Thus, it seamlessly unifies both modes operating with and without the enhancement bit-stream.

Our neural-network-based approach is named geometric residual analysis and synthesis for PCC, or GRASP-Net for short.
The main contributions of our GRASP-Net are:
\begin{enumerate}
\item We propose a heterogeneous PCC framework with different processing backbones to tackle the sparse local details of point clouds. Especially, a point-based network is employed for converting the local details to latent features that are more densely populated; then a sparse CNN is applied to remove the spatial redundancy in the features, leading to an enhancement bit-stream representing the geometric details.
\item Our proposal operates properly when the enhancement bit-stream is unavailable. It is achieved by seamlessly switching to a skip mode which generates at the receiver side the geometric details solely from a coarse point cloud.
\item Experimentation shows that the proposed GRASP-Net provides state-of-the-art compression performance on both dense and sparse point clouds.
\end{enumerate}

\vspace{-10pt}
\section{Related Works}
\label{sec:related}
State-of-the-art non-learning-based PCC methods include MPEG G-PCC, MPEG V-PCC, and Draco, see \cite{graziosi2020overview} for an overview.
Recent advances in deep learning provide opportunities to outperform them~\cite{quach2022survey}.
Deep-learning-based approaches purely based on voxels or octrees attempt to organize 3D point sets in the first place, putting the representability of geometric details at risk.
Differently, native learning approaches keep the local geometry untouched but are faced with the irregularity of 3D point points.

\textbf{Voxel-/Octree- based PCC}:
Based on the seminal works of the factorized prior model~\cite{balle2016end} and the scale hyperprior model~\cite{balle2018variational} for end-to-end image compression, Quach~\etal~\cite{quach2019learning,quach2020improved} apply 3D convolutional layers to compress voxelized point cloud geometry in a lossy manner.
A subsequent work, ADL-PCC~\cite{guarda2020adaptive}, chooses different network models according to the local contents; while PCGCv1~\cite{wang2021lossy} proposes an advanced cross entropy loss for better training.
These approaches relying on regular 3D convolutions are memory-inefficient since most voxels are empty in a real-world point cloud.
Differently, PCGCv2~\cite{wang2021multiscale} applies 3D sparse convolutions to estimate voxel occupancy, which only operates on non-empty voxels.
Essentially, voxel-based methods convert 3D point clouds into regular Euclidean data.
As elaborated in Section~\ref{sec:intro}, their effectiveness is harmed by the local sparsity of point clouds, especially when operating on point clouds with high bit-depths.

Building on top of the voxelized point clouds, the octree representation is widely used for lossless PCC.
It is achieved by an adaptive context model predicting the voxel occupancy status.
Representative works along this thread include OctSqueeze~\cite{huang2020octsqueeze}, VoxelContext-Net~ \cite{que2021voxelcontext}.
Inspired by PCGCv2~\cite{wang2021multiscale}, the recently proposed SparsePCGC~\cite{wang2021sparse} applies sparse CNN for octree coding.

\textbf{Native learning of point cloud}:
Emerging techniques for learning on native point clouds relieve the inherent problem of voxelization.
A point-based encoder (analysis network) directly operates on a point set to obtain a latent feature or codeword which abstracts the input point set.
The pioneering work, PointNet~\cite{qi2017pointnet}, combines multi-layer perceptron (MLP) and pooling operation for encoding with point permutation invariant.
Subsequent proposals, \eg, PointCNN~\cite{li2018pointcnn}, PointConv~\cite{wu2019pointconv}, and KPConv~\cite{thomas2019kpconv}, improves PointNet by counting the relationship between neighboring points.
As opposed to an encoder, a point-based decoder (synthesis network) reconstructs a point set from a latent feature.
The work Latent~GAN~\cite{achlioptas2018learning} directly generates a point cloud by a series of MLP layers; while other decoders such as FoldingNet~\cite{yang2018foldingnet}, AtlasNet~\cite{groueix2018papier}, 3D Point Capsule Net~\cite{zhao20193d}, and TearingNet~\cite{pang2021tearingnet} embed a global topology for reconstruction.
The Latent GAN decoder imposes no constraints on the point clouds.
Therefore, it is suitable for generating intricate local structures, as verified by Fan~\etal~\cite{fan2017point}.

\textbf{Point-based PCC}:
There exists a few PCC approaches with point-based neural networks.
Many of them, such as \cite{yan2019deep}, \cite{huang20193d}, and \cite{you2021patch}, fail to justify their proposals on real-world point cloud data collected from depth sensors, \eg, the test sequence recommended by the MPEG group~\cite{ctcgpcc}.
In contrast, the proposed GRASP-Net, to our best knowledge, is the \emph{first} work utilizing point-based learning to achieve promising compression on both real-world dense and sparse point clouds.
A recent work DEPOCO~\cite{wiesmann2021deep} compresses dense point cloud maps by employing KPConv~\cite{thomas2019kpconv} for local feature extraction.
However, their extracted features are directly arithmetically encoded; while we further exploit the spatial correlation among neighboring features for additional compression.

\section{Heterogeneous Framework for PCC}
\label{sec:framework}
We first provide an analysis of the local sparsity of 3D point clouds by inspecting their dimensions using fractal geometry~\cite{falconer2004fractal}.
Then our proposed heterogeneous architecture for PCC is presented, followed by a discussion on its skip mode.

\subsection{Local Sparsity Analysis}\label{ssec:sparsity_analysis}
\textbf{Background}:
Point clouds approximate object/scene surfaces with 3D points.
These 3D points are of limited precision, reflected by their bit-depth values.
For instance, the $x$, $y$, and $z$ coordinates of a typical $10$-bit point cloud are integers ranging from $0$ to $1023$.
It can be quantized to a coarser spatial scale, by removing the most significant bit of all coordinates followed by merging the duplicate points, leading to a $9$-bit point cloud with coordinates ranging from $0$ to $511$.
Suppose the original $10$-bit point cloud is very dense, then its point number is approximately $8$ times the 9-bit version since an integer position in the $9$-bit space corresponds to $8$ different positions in the $10$-bit space.
On the other hand, if this point cloud is very sparse, then the number of points stays roughly the same in both the 10-bit and the 9-bit versions because an isolated point wouldn't be merged after quantization.
Thus, the increment factor of the point number from bit-depth $n-1$ to $n$, denoted as $r_n$, reflects the 
\emph{sparsity} level of a point cloud at the $n$-th spatial scale.

According to fractal analysis~\cite{falconer2004fractal}, the logarithm of $r_n$---computed as $D_n=\log_{2}(r_n)$, converges to the \emph{fractal dimension} as the scale/bit-depth $n$ increases.
Intuitively, the fractal dimension represents a shape's complexity by measuring its space-filling capability.
It does not have to be an integer for general geometry like fractals.
However, its value \emph{coincides} with regular dimension by definition, \eg, the fractal dimension computes as $2$ for 2D surfaces, and $1$ for 1D curves.
This also holds true for the approximate fractal dimension $D_n$~\cite{falconer2004fractal}, which measures the ``effective'' dimension of a point cloud at the $n$-th scale.
For example, a point cloud with $r_n=8$ or $D_n=3$ implies that at the $n$-th scale, it appears like a solid 3D volume.

\textbf{Dimensionality analysis}:
We analyze the sparsity of real-world point clouds by checking how their approximate fractal dimensions $D_n$ change as the scale/bit-depth $n$ increases. 
Given a point cloud, we plot its $D_n$ values at different scales $n$, resulting in a plot that we call a \emph{dimension spectrum}.
A few representative 12-bit dense point clouds and 18-bit sparse LiDAR sweeps from the test sequences of MPEG~\cite{ctcgpcc} are chosen in this study.

Fig.~\ref{fig:dim} shows their dimension spectrums.
At its bottom, two point clouds, ``house\_without\_roof'' (dense) and the first frame of ``Ford\_01\_q\_1mm'' (sparse), are visualized at different spatial scales.
From Fig.~\ref{fig:dim_dense}, we see that at most scales, the dimensions of the dense point clouds are close to $2$, meaning that they managed to delineate object surfaces that are 2D (Fig.~\ref{fig:dim_dense}-(ii)).
Differently, the dimensions of LiDAR sweeps (From Fig.~\ref{fig:dim_sparse}) are around $1$ in many cases.
It is due to the line patterns~(Fig.~\ref{fig:dim_sparse}-(ii)) introduced by spinning LiDARs~\cite{li2020lidar}.
However, a common trend for both cases is that their dimensions $D_n$ gradually decrease and approach $0$ at the finest scales.
It indicates that as granularity increases, their geometry degenerate to isolated points (Fig.~\ref{fig:dim_dense}-(iii) and Fig.~\ref{fig:dim_sparse}-(iii)).

Thus, real-world point clouds behave differently at different scales---they resemble continuous geometry at coarser scales but are locally sparse and disconnected.
To achieve effective compression (or other processing tasks), a heterogeneous approach combining different methodologies at different scales would be beneficial.

\subsection{Heterogeneous Architecture}\label{ssec:arch}
\begin{figure}
  \centering
  \includegraphics[width=0.95\columnwidth]{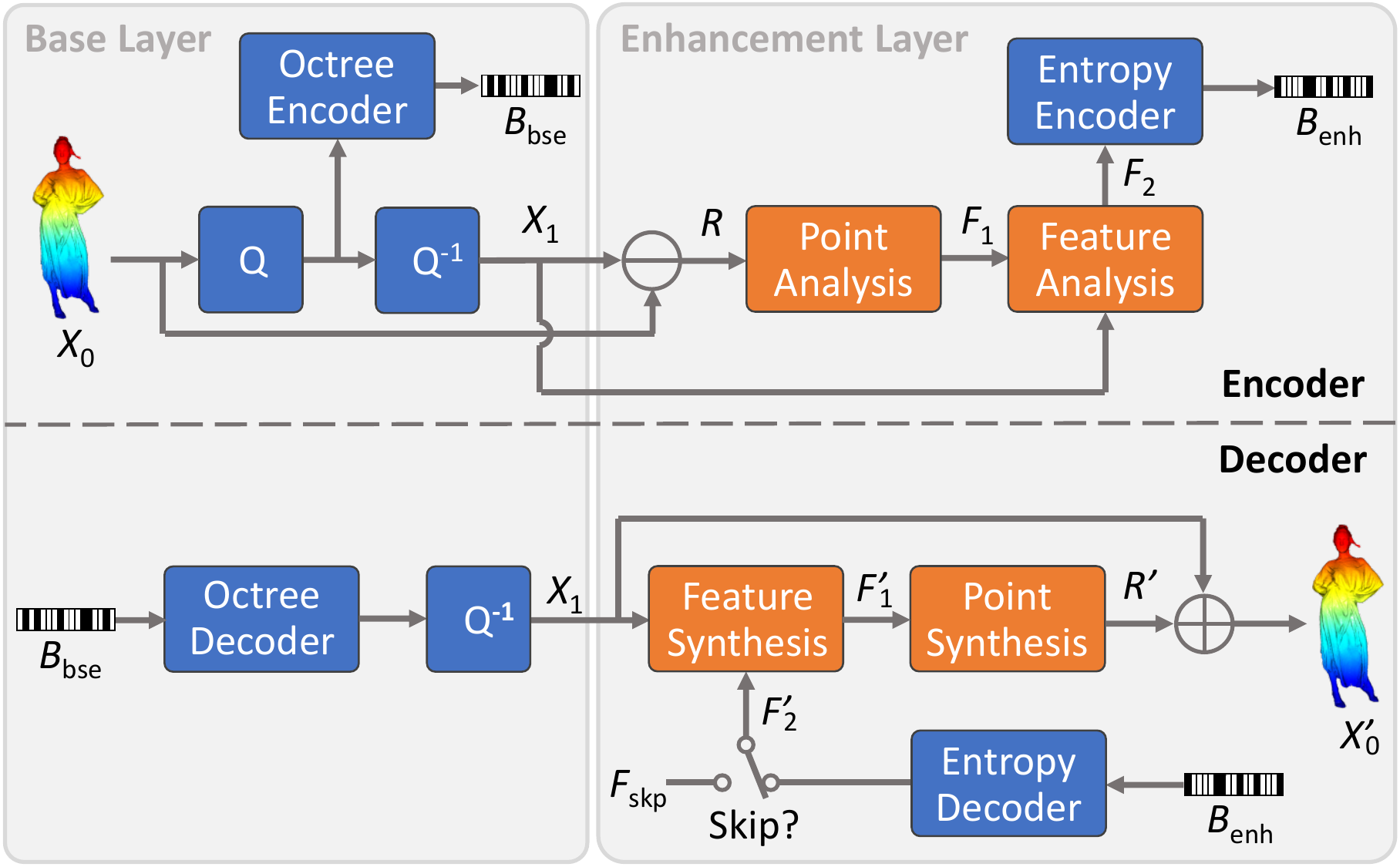}
  \caption{ Architecture of our GRASP-Net. Q is quantization while Q$^{-1}$ is dequantization. The orange blocks are neural networks, where the Point Analysis/Synthesis Networks operate on raw 3D points while the Feature Analysis/Synthesis Networks are sparse CNNs operate on voxels.}
  \label{fig:framework}
  \vspace{-15pt}
\end{figure}

We propose a heterogeneous architecture for PCC: \emph{octree-/voxel-based} methods are employed when operating with the coarser scales; while \emph{point-based} learning is applied to digest the finest scales.
Fig.~\ref{fig:framework} presents the overall architecture of the proposed GRASP-Net.
Both the encoder and decoder consist of a base layer for the compression of a coarse version of the input, and an enhancement layer to additionally compress the local geometric details.

\textbf{Encoder}:
Given an input point cloud $X_0$, the base layer obtains its coarse version $X_1$ by quantization and dequantization (Q and Q$^{-1}$ in Fig.~\ref{fig:framework}).
The quantization module divides the 3D coordinates of $X_0$ by a step size $s>1$ followed by rounding, then the duplicate points are merged; while the dequantization module simply multiplies the coordinates by $s$.
See an illustrative example in Fig.~\ref{fig:input} and b.
Intuitively, $X_1$ is obtained by removing the last $\log_{2}{s}$ bits from $X_0$.
Right after the quantization module, the obtained point cloud is compressed losslessly as a bitstream $B_\textrm{bse}$ by an octree encoder (\eg, G-PCC octree~\cite{graziosi2020overview}).
Since $B_\textrm{bse}$ contains all the information of the coarse point cloud $X_1$, the enhancement layer further encodes the last $\log_{2}{s}$ bits of $X_0$.

In the enhancement layer, we first feed $X_0$ and its coarse version $X_1$ to a \emph{geometric subtraction module} ($\ominus$ in Fig.~\ref{fig:framework}).
It subtracts $X_1$ from $X_0$ to obtain the geometric residual $R$ which are native point sets containing the local details of $X_0$.
We will elaborate on the geometric subtraction module (as well as a geometric addition module on the decoder side) in Section~\ref{ssec:point}.
Next, the \emph{Point Analysis Network} is applied to the residual $R$.
It is a point-based encoder network that generates for each point in $X_1$ a latent feature.
For example, a 3D point $A$ in $X_1$ will be equipped with a feature vector $\mathbf{f}_A$, which abstracts the local details in $X_0$ that are close to $A$.

Based on the voxelized geometry of $X_1$, the set of pointwise features (represented by $F_1$) is gradually down-sampled with a sparse CNN---the \emph{Feature Analysis Network}.
It leverages $X_1$, which has better continuity than $X_0$, to further discover the redundancy between the neighboring features in $F_1$.
The set of down-sampled features, denoted as $F_2$, is finally entropy encoded as an enhancement bit-stream $B_\textrm{enh}$, dedicated to representing local details.
Note that our entropy encoder (and decoder) for the compression of $F_2$ is based on the factorized prior model~\cite{balle2018variational}.
In this step, an extra quantization operation is applied to $F_2$, which is absorbed within the entropy encoder.
Thus, the compression of $F_2$ is lossy.

\textbf{Decoder}:
Having received the bit-streams $B_\textrm{bse}$ and $B_\textrm{enh}$, the base layer decodes $B_\textrm{bse}$ followed by a dequantization (with step size $s$) to obtain the coarse point cloud $X_1$.

In the enhancement layer, the bit-stream $B_\textrm{enh}$ is firstly entropy decoded to obtain the down-sampled feature set, $F'_2$---it is the quantized version of $F_2$ on the encoder.
Then based on the voxelized geometry of the coarse point cloud $X_1$, $F'_2$ is gradually up-sampled by the \emph{Feature Synthesis Network}---a sparse CNN with deconvolutional layers.
The upsampled feature set, denoted as $F'_1$, is then passed to the \emph{Point Synthesis Network} using MLP layers for recovering the residual component $R'$.
Then the summation module ($\oplus$ in Fig.~\ref{fig:framework}) adds back the residual $R'$ to the coarse point cloud $X_1$ through a residual connection~\cite{he2016deep}, leading to the reconstructed point cloud $X'_0$ with fine details.

\subsection{Skip Mode}
In case of packet loss or high latency due to network congestion, the enhancement bit-stream $B_\textrm{enh}$ may not be available to the decoder. 
Our encoder may also intentionally skip the coding of the enhancement bit-stream.
Under such circumstances, the decoder is still expected to operate properly. 
Thus, we propose to accommodate the skip mode which reconstructs $X'_0$ based on $B_\textrm{bse}$ alone.

As shown in Fig.~\ref{fig:framework}, a \emph{skip flag} indicates whether the skip mode is triggered.
When skip equals false, the decoder operates as presented in Section~\ref{ssec:arch}.
However, when the skip flag is switched to true, $B_\textrm{enh}$ is no longer required by the enhancement layer.
Instead, we let $F_2'$ be a constant feature set $F_\textrm{skp}$, \eg, all entries filled with $1$s for feeding to the Feature Synthesis Network.
In this case, the two Synthesis Networks generate the residual $R'$ with the geometry of the coarse point cloud $X_1$ alone, akin to post-processing on $X_1$ with super-resolution~\cite{borges2022fractional}.
From Fig.~\ref{fig:framework}, no matter whether the skip mode is switched on or not, our network stays unchanged---it unifies both cases with or without the enhancement bit-stream.


\section{Geometric Residual Coding}
\label{sec:residual}
This section illustrates how the geometric details---the last $\log_{2}s$ bits of $X_0$---are compressed heterogeneously by the enhancement layer (Fig.~\ref{fig:framework}).

\subsection{Point-based Residual Learning}\label{ssec:point}
\begin{figure}
  \centering
  \includegraphics[width=1\columnwidth]{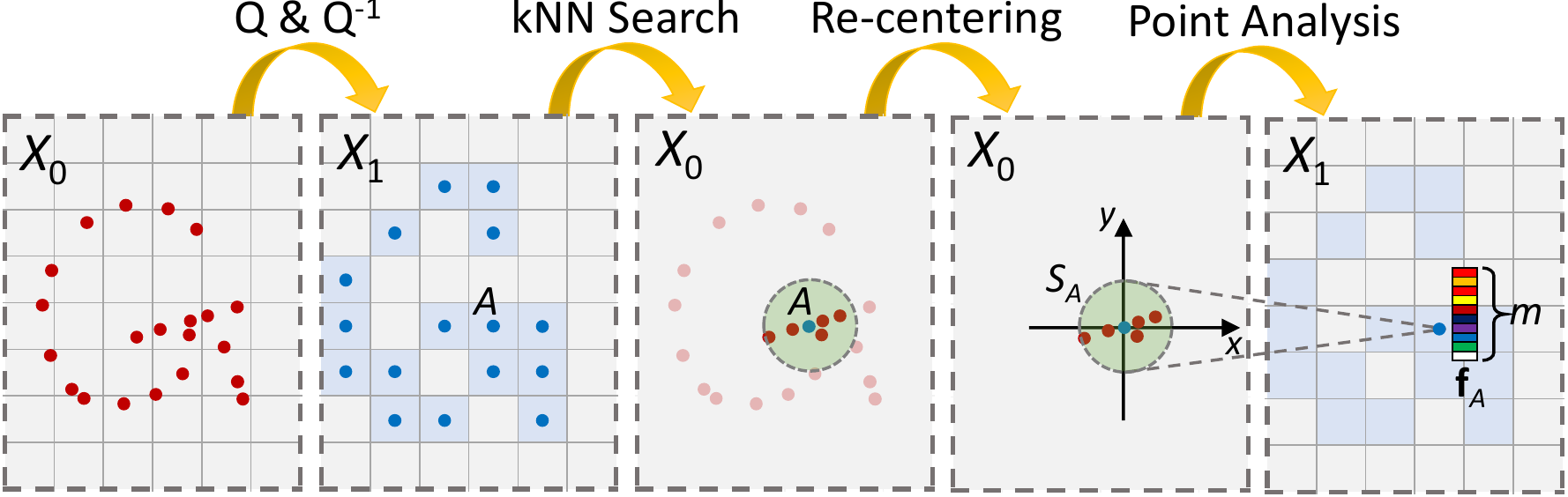}\\
  \vspace{-14pt}
  \subfigure[]{\includegraphics[width=0.17\columnwidth]{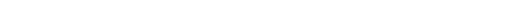}\label{fig:input}}
  \subfigure[]{\includegraphics[width=0.2\columnwidth]{figures/place_holder.png}\label{fig:coarse}}
  \subfigure[]{\includegraphics[width=0.2\columnwidth]{figures/place_holder.png}\label{fig:grouping}}
  \subfigure[]{\includegraphics[width=0.2\columnwidth]{figures/place_holder.png}\label{fig:recentering}}
  \subfigure[]{\includegraphics[width=0.15\columnwidth]{figures/place_holder.png}\label{fig:pointnet}}
  \caption{Geometric subtraction and analysis on native points. For any point $A\in X_1$, the geometric subtraction module applies kNN search (b$\rightarrow$c) and re-centering (c$\rightarrow$d) to obtain a residual point set $S_A$. It is then fed to the Point Analysis Network to obtain a latent feature $\textbf{f}_A$ (d$\rightarrow$e).}
  \label{fig:point_learning}
  \vspace{-5pt}
\end{figure}
A point-based encoder network is used to digest the geometric residual/details of $X_0$ during encoding.
With a residual connection, our decoder learns to generate the geometric residual by a point-based decoder network.

\textbf{Geometric subtraction $\ominus$}:
The enhancement layer begins with ``subtracting'' $X_1$ from the input point cloud $X_0$ for extracting the geometric residual $R$.
For a coarse point cloud $X_1$ with $N$ points, we define $R$ as the collection of $N$ \emph{residual point sets} where each of them corresponds to a point in $X_1$.
Given a point $A\in X_1$, its residual point set, denoted as $S_A$, is obtained in two steps.
Firstly, we retrieve the local details of $X_0$ that are close to $A$ with a k-nearest-neighbor (kNN) search, which looks for the $k$ closet points of $A$ in $X_0$ (Fig.~\ref{fig:grouping}).
Secondly, these $k$ neighboring points are re-centered to a local coordinate system relative to $A$ by coordinate-wise subtraction (Fig.~\ref{fig:recentering}).
Particularly, a neighboring point $(x, y, z)$ is translated to $(x-x_A, y-y_A,z-z_A)$ where $(x_A, y_A, z_A)$ denote the coordinates of $A$.
The $k$ re-centered neighbors form the residual point set $S_A$.
Having obtained all the $N$ residual point sets, they are grouped as the residual $R$ and output by the geometric subtraction module.

\textbf{Point analysis}:
Next, a shared Point Analysis Network is applied to every residual point set in $R$, leading to $N$ latent features.
An example is shown in Fig.~\ref{fig:pointnet} where the residual point set $S_A$ is abstracted as a feature vector $\textbf{f}_A\in\mathbb{R}^m$.
The obtained feature vectors, associated with all the points of $X_1$, are grouped as a feature set $F_1$ (Fig.~\ref{fig:framework}).
It is a high-level abstraction of the geometric residual $R$.
The PointNet~\cite{qi2017pointnet} architecture is chosen as our Point Analysis Network for its simplicity and effectiveness, though more complicated point-based encoders, as discussed in Section~\ref{sec:related}, can also be applied.
Note that the geometric subtraction, together with the point analysis, resemble the Set Abstraction layer presented in PointNet++\cite{qi2017pointnet++}.
Differently, our design manages to transform the unevenly and sparsely distributed local details in $X_0$ as more regularly-distributed features residing on $X_1$, making it suitable to be further processed by convolutional layers.

\textbf{Point synthesis}:
On the decoder side, the Point Synthesis Network reverses the operation of the Point Analysis Network.
It is a shared network that applies to every feature vector in the decoded feature set $F_1'$, converting $F_1'$ to the decoded geometric residual $R'$ (decoder of Fig.~\ref{fig:framework}).
Given a feature vector $\textbf{f}_A'$ associated with the point $A$ in $X_1$, the Point Synthesis Network generates a decoded residual point set, denoted as $S_A'$.
Then the collection of all the decoded residual point sets forms the decoded geometric residual $R'$.
In our work, we choose the Latent~GAN decoder~\cite{achlioptas2018learning} to implement our Point Synthesis Network for its simplicity and capability of generating intricate geometric details, as discussed in Section~\ref{sec:related}.

\textbf{Geometric addition $\oplus$}:
With a residual connection that introduces the coarse point cloud $X_1$ to the geometric addition module (decoder of Fig.~\ref{fig:framework}), the decoded geometric residual $R'$ is added back to $X_1$, leading to the decoded point cloud $X_0'$. 
The geometric addition reverses the operation of geometric subtraction.
It translates each residual point set of $R'$ by the corresponding point in $X_1$.
Specifically, given a decoded residual point set $S_A'$, the geometric addition module adds the coordinates of $A$ to all the points in $S_A'$, \ie, a point $(x, y, z)\in S_A'$ becomes $(x+x_A, y+y_A, z+z_A)$.
Having shifted all the residual point sets of $R'$, the shifted points are finally aggregated as the decoded point cloud $X_0'$.

\subsection{Voxelized Feature Coding}\label{ssec:voxel}
\begin{figure}
  \centering
  \includegraphics[width=1\columnwidth]{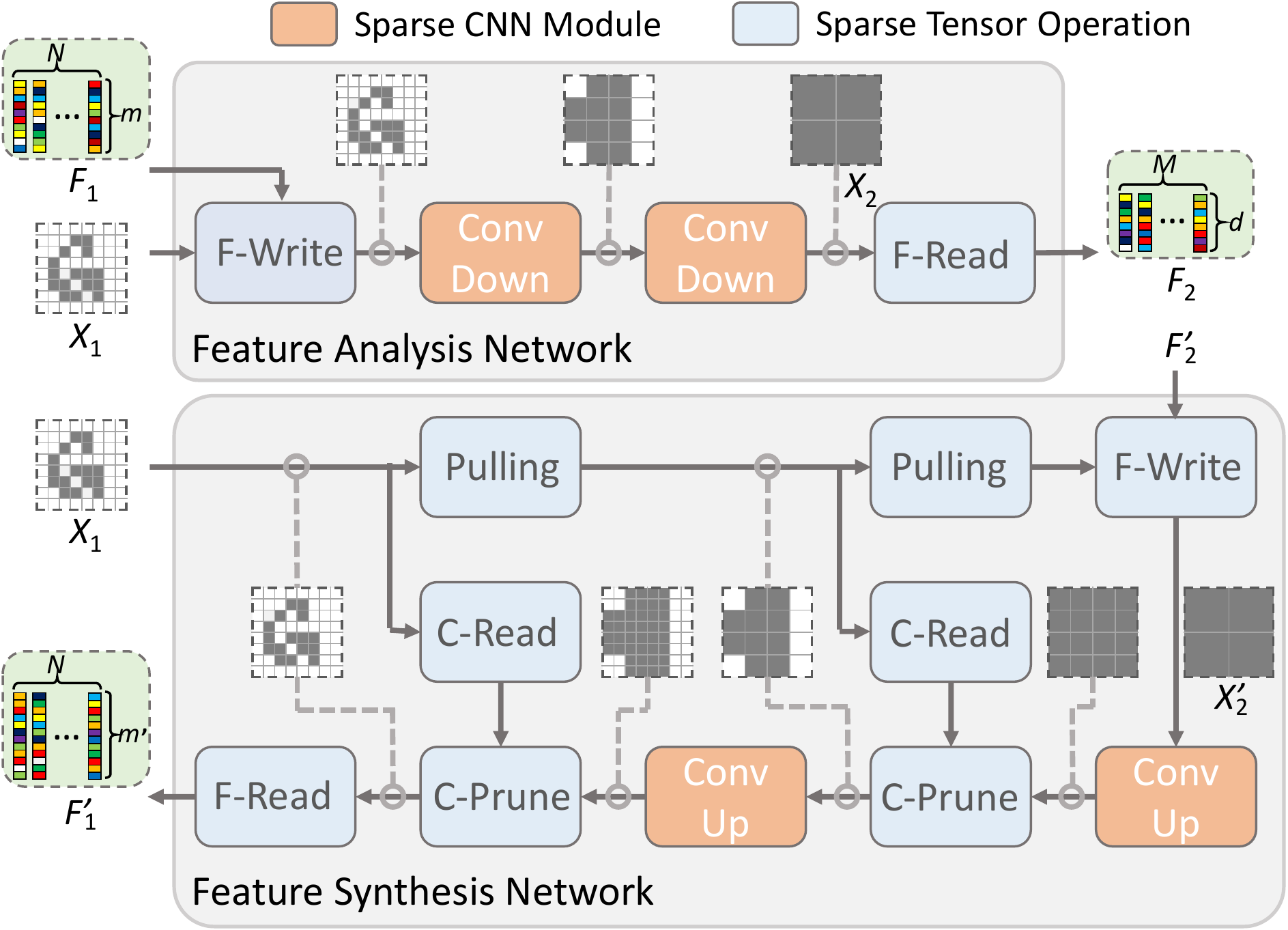}
  \caption{Geometric feature analysis and synthesis with voxel-based representation. The Conv Down module performs down-sampling to a sparse tensor while the Conv Up module performs up-sampling. They are built by sparse convolutional (and deconvolutional) layers.}
  \label{fig:voxel_learning}
  \vspace{-5pt}
\end{figure}

In this section, we explore the correlation between neighboring latent features in the feature set $F_1$ for compression purpose.
Since the coarse point cloud $X_1$ can be naturally represented by voxel grids with length $s$ (\eg, Fig.~\ref{fig:coarse}), we propose to use two sparse CNNs (as shown in Fig.~\ref{fig:voxel_learning}) to analyze and synthesize the feature vectors in $F_1$, respectively.

Sparse 3D convolutional layers operate on \emph{sparse 3D tensors}.
Different from regular 3D tensors, a sparse tensor maintains two components: (i)~a list of occupied 3D voxels (represented by their voxel coordinates); and (ii) and a list of features where each feature is associated with an occupied voxel.
Thus, a sparse tensor only requires memory for the occupied voxels.
When a sparse convolution is applied, the convolution will only take place on the occupied voxels, making it highly efficient compared to regular convolution.
In the sequel, the point cloud $X_1$ is viewed as a 3D sparse tensor, where we let the scalar $1$ be a stuffing feature associated with every occupied voxel of $X_1$.

\textbf{Feature analysis}:
On the encoder side, the feature set $F_1$ is gradually down-sampled by the Feature Analysis Network, as shown by the upper part of Fig.~\ref{fig:voxel_learning}.
To begin with, the \emph{Feature-Writer} module (F-Write in Fig.~\ref{fig:voxel_learning}) replaces all the stuffing features in $X_1$ by the corresponding feature vectors in $F_1$, \eg, a latent feature $\textbf{f}_A$ is assigned to the voxel representing point $A$.
The resulting sparse tensor is then down-sampled $4$ times by two \emph{Conv Down} modules.
A Conv Down module includes a sparse convolutional layer with stride $2$ for down-sampling by a factor of $2$, followed by Inception ResNet blocks (IRN)~\cite{szegedy2017inception,wang2021multiscale} for efficient feature learning.
The output tensor, $X_2$, is then passed to a \emph{Feature-Reader} module (F-Read in Fig.~\ref{fig:voxel_learning}) to read its feature vectors.
Suppose the number of output channels of the second Conv Down module is $d$, then the feature vectors also have a length $d$.
These feature vectors are grouped as the down-sampled feature set $F_2$.
They are then encoded as the enhancement bit-stream $B_\textrm{enh}$ by the entropy encoder in Fig.~\ref{fig:framework}.

\textbf{Feature synthesis}:
On the decoder side, the Feature Synthesis Network up-samples $F_2'$ by utilizing the geometry of $X_1$.
To facilitate the up-sampling, it additionally includes modules to down-sample $X_1$ in the first place.
Overall, it takes a U-Net architecture with skip connections, as shown in the lower part of Fig.~\ref{fig:voxel_learning}.

Particularly, $X_1$ (as a sparse tensor) is first down-sampled consecutively by two \emph{pulling} layers (Pulling in Fig.~\ref{fig:voxel_learning}), where each of them reduces the size of the input tensor by a factor of two.
The purpose of the pulling layers is to reproduce the occupied voxels of the tensor $X_2$.
Then the obtained tensor (output by the second pulling layer) is fed to a Feature-Writer module, which replaces its stuffing features by $F_2'$, resulting in the sparse tensor $X_2'$.
We note that $X_2$ and $X_2'$ have the \emph{same geometry}---their occupied voxels are the same, though they have different features, $F_2$ and $F_2'$, respectively.
When the skip mode is turned off, $F_2'$ is the quantized version of $F_2$; otherwise, it is the predefined feature set $F_\textrm{skp}$.

$X_2'$ is then up-sampled by two \emph{Conv Up} modules.
A Conv Up module contains a deconvolutional layer with stride $2$ for up-sampling with a factor of $2$, followed by IRN blocks for feature aggregation.
Right after each Conv Up module, we remove the redundant occupied voxels based on the geometry of $X_1$ (or the down-sampled version of $X_1$).
For example, after the second Conv Up module, a \emph{Coordinate-Pruning} module (C-Prune in Fig.~\ref{fig:voxel_learning}) will remove the occupied voxels for which the \emph{Coordinate-Reader} module (C-Read in Fig.~\ref{fig:voxel_learning}) did not find/provide by reading the voxel coordinates of $X_1$.
While the PCGCv2 decoder~\cite{wang2021multiscale} prunes voxels based on the classification results, our solution fully exploits the geometry of $X_1$ for pruning, resulting in a refined up-sampling result.

\begin{figure*}[htbp]
  \centering \scriptsize 
  \includegraphics[width=1.5\columnwidth]{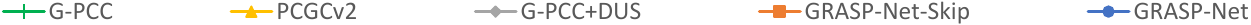}\\
  \subfigure{\includegraphics[height=120pt]{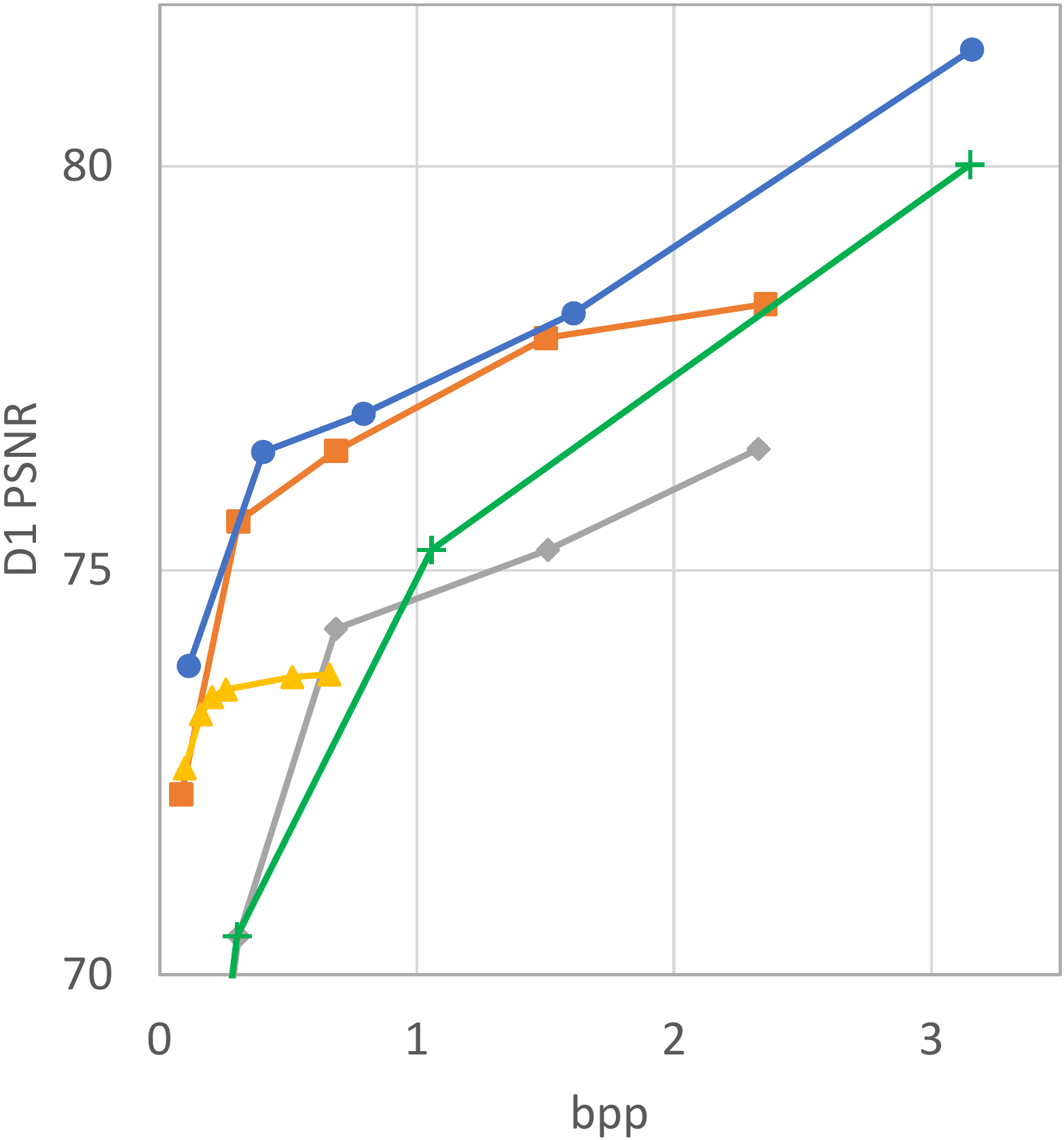}}\hspace{5pt}
  \subfigure{\includegraphics[height=120pt]{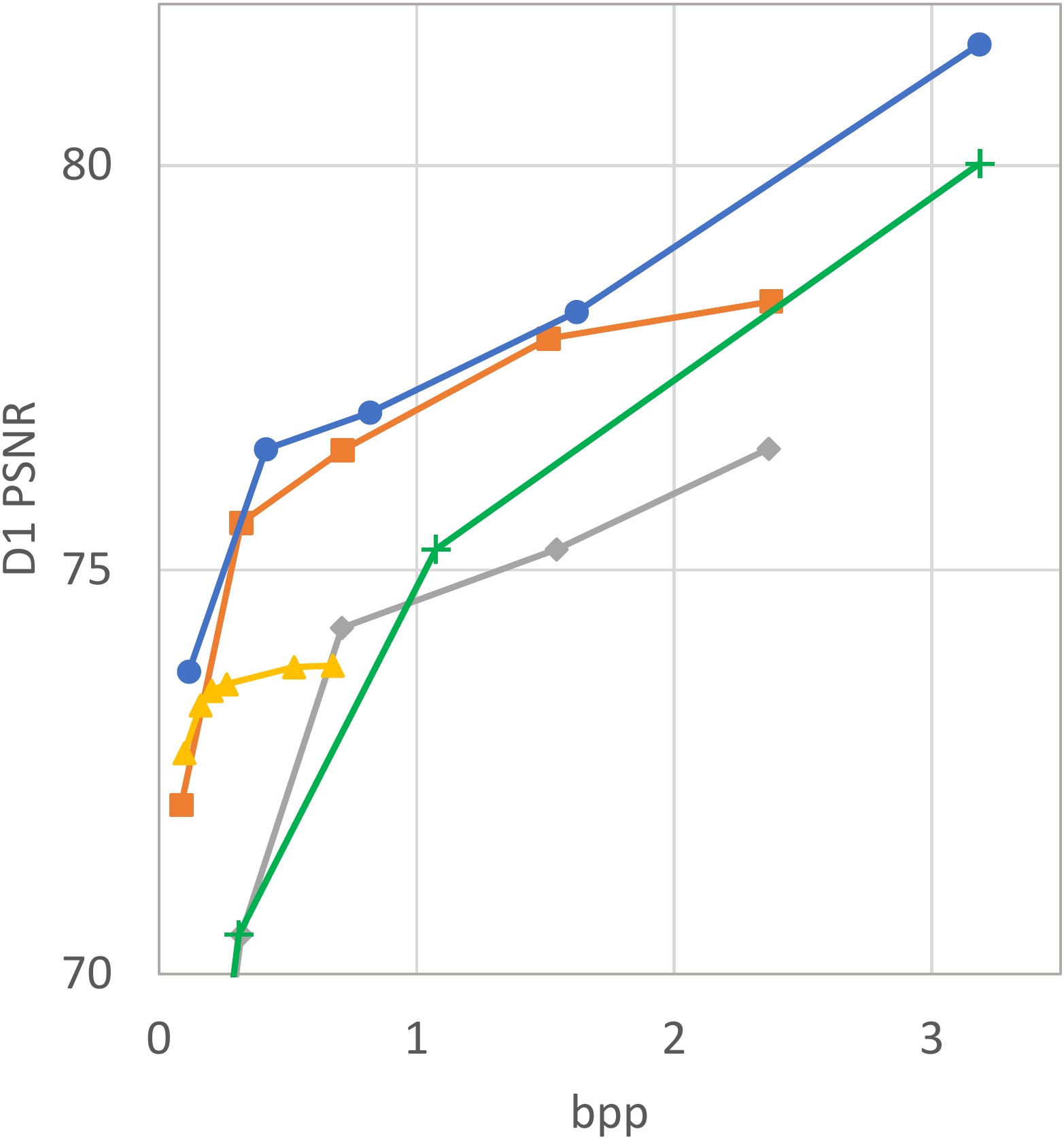}}\hspace{5pt}
  \subfigure{\includegraphics[height=120pt]{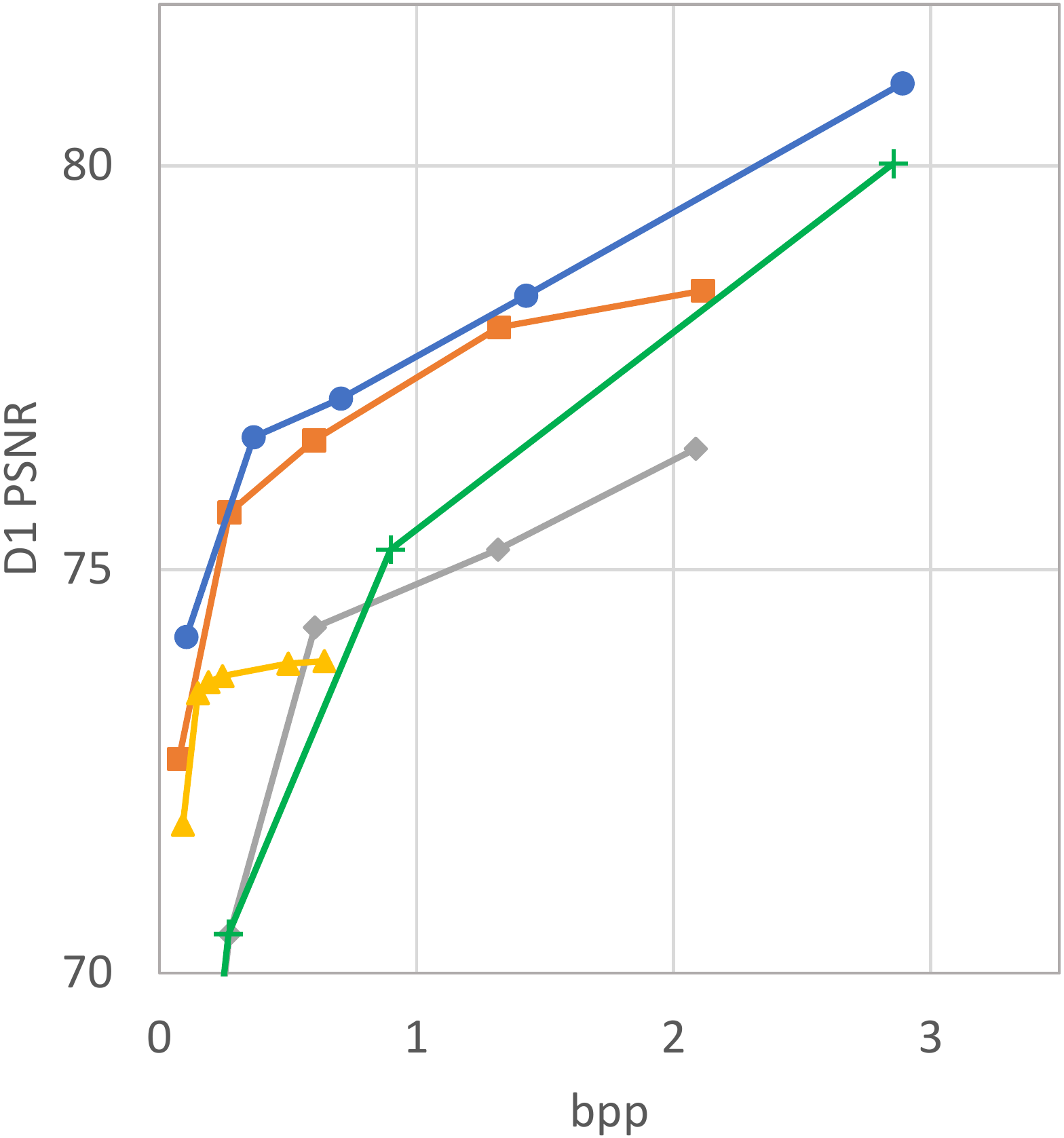}}\hspace{5pt}
  \subfigure{\includegraphics[height=120pt]{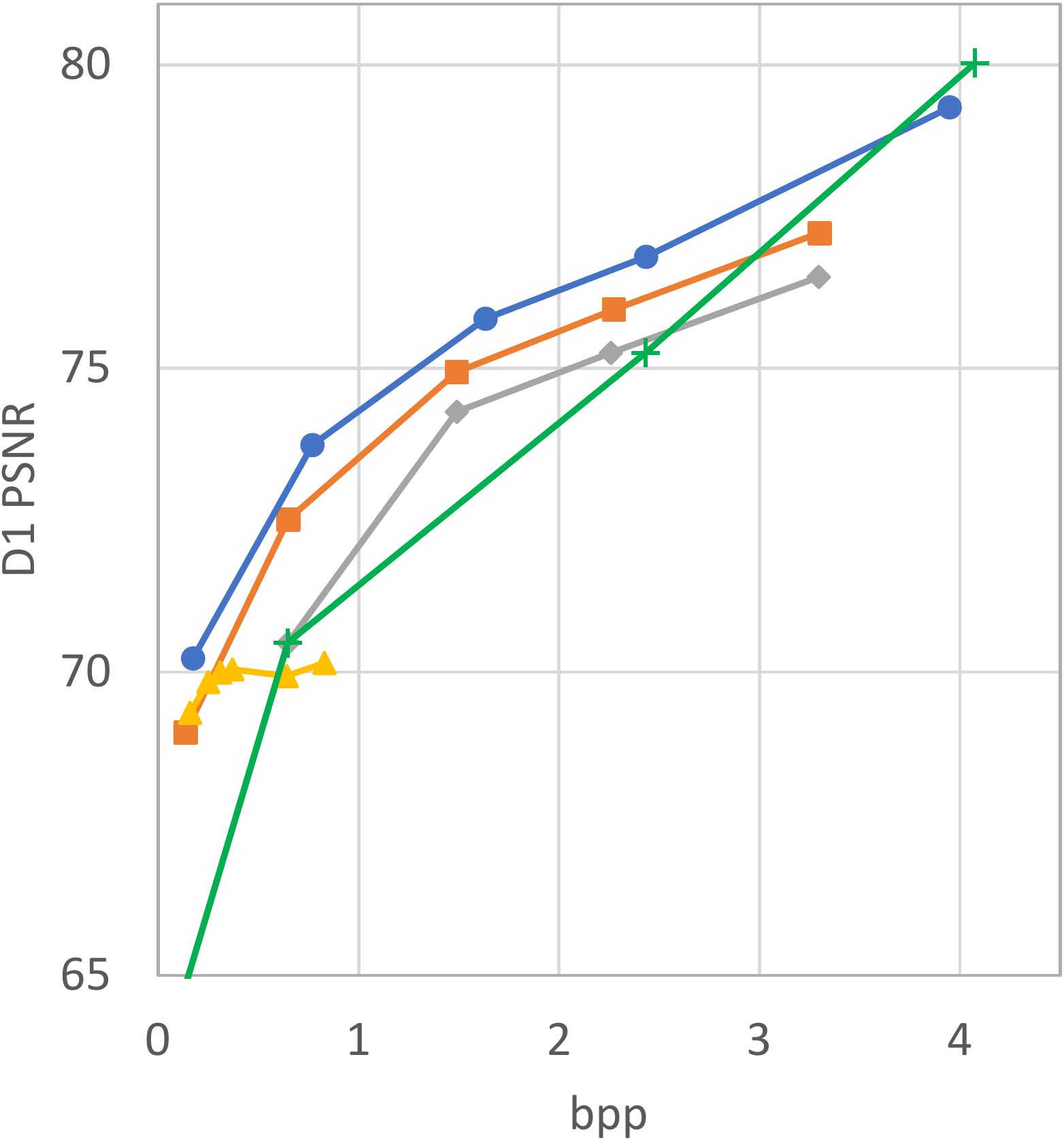}}\\\vspace{-10pt}
  \setcounter{subfigure}{0}
  \subfigure[``longdress\_viewdep\_vox12'']{\includegraphics[height=120pt]{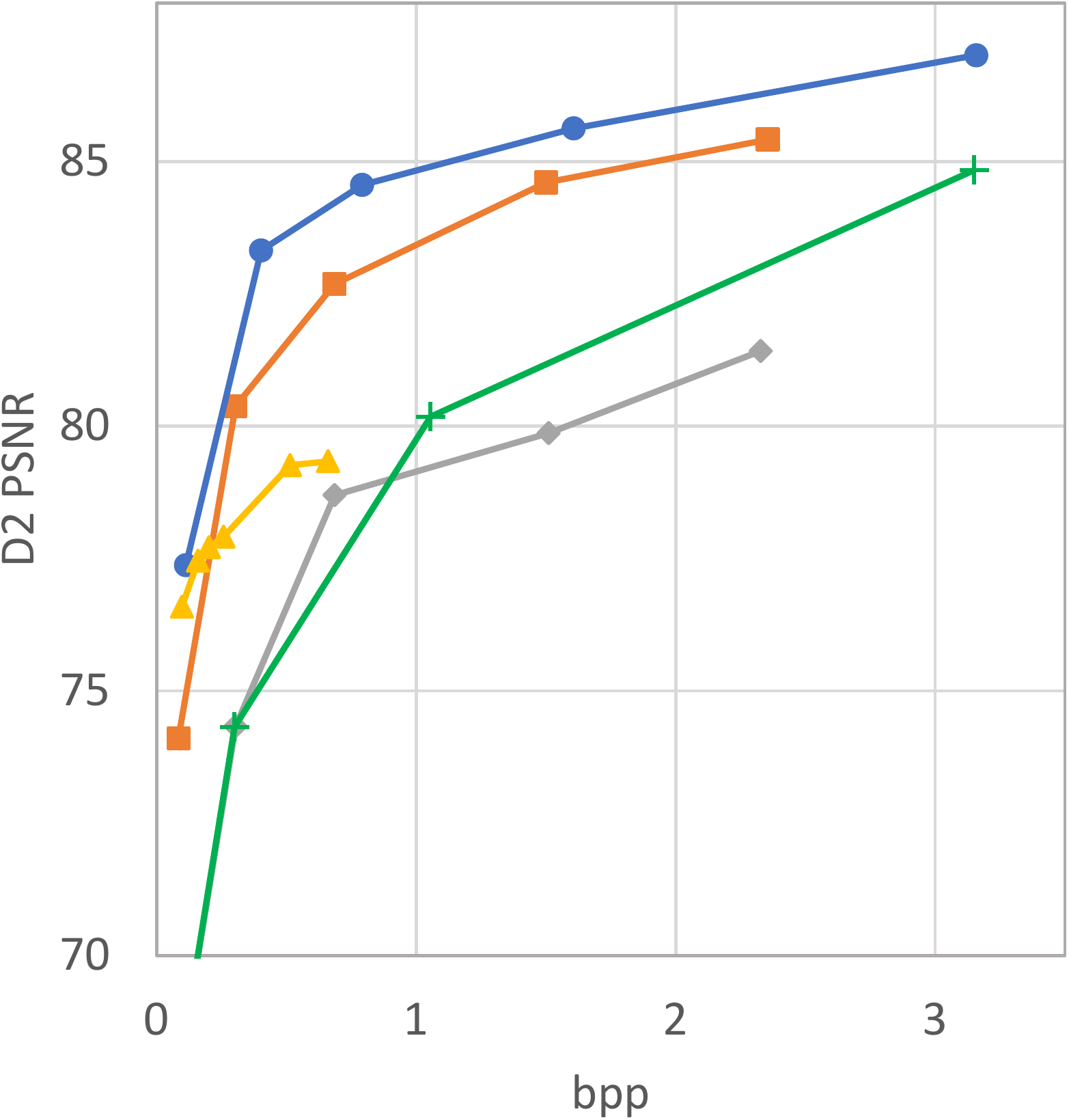}\label{fig:rd_longdress}}\hspace{5pt}
  \subfigure[``soldier\_viewdep\_vox12'']{\includegraphics[height=120pt]{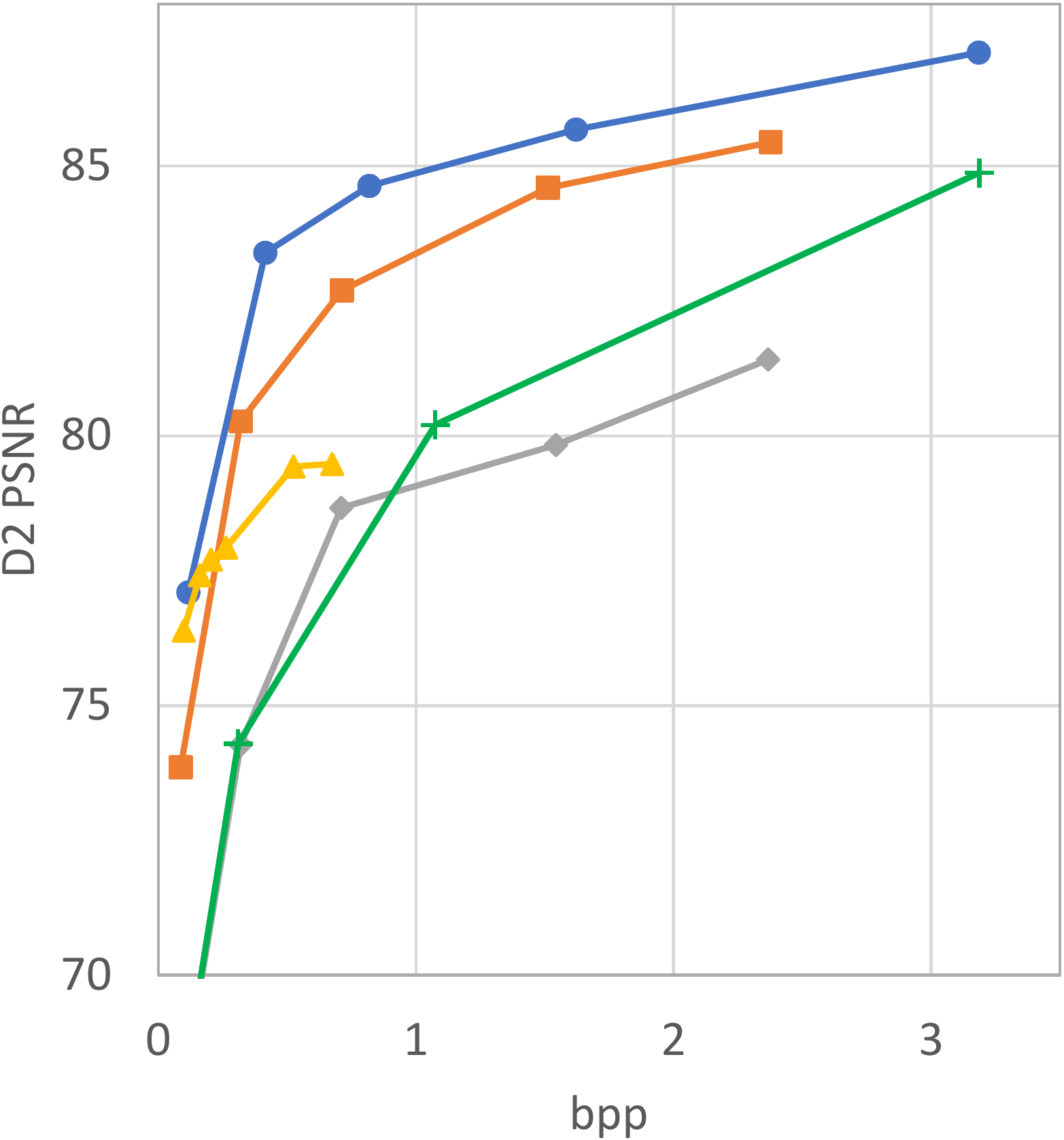}\label{fig:rd_soldier}}\hspace{5pt}
  \subfigure[``boxer\_viewdep\_vox12'']{\includegraphics[height=120pt]{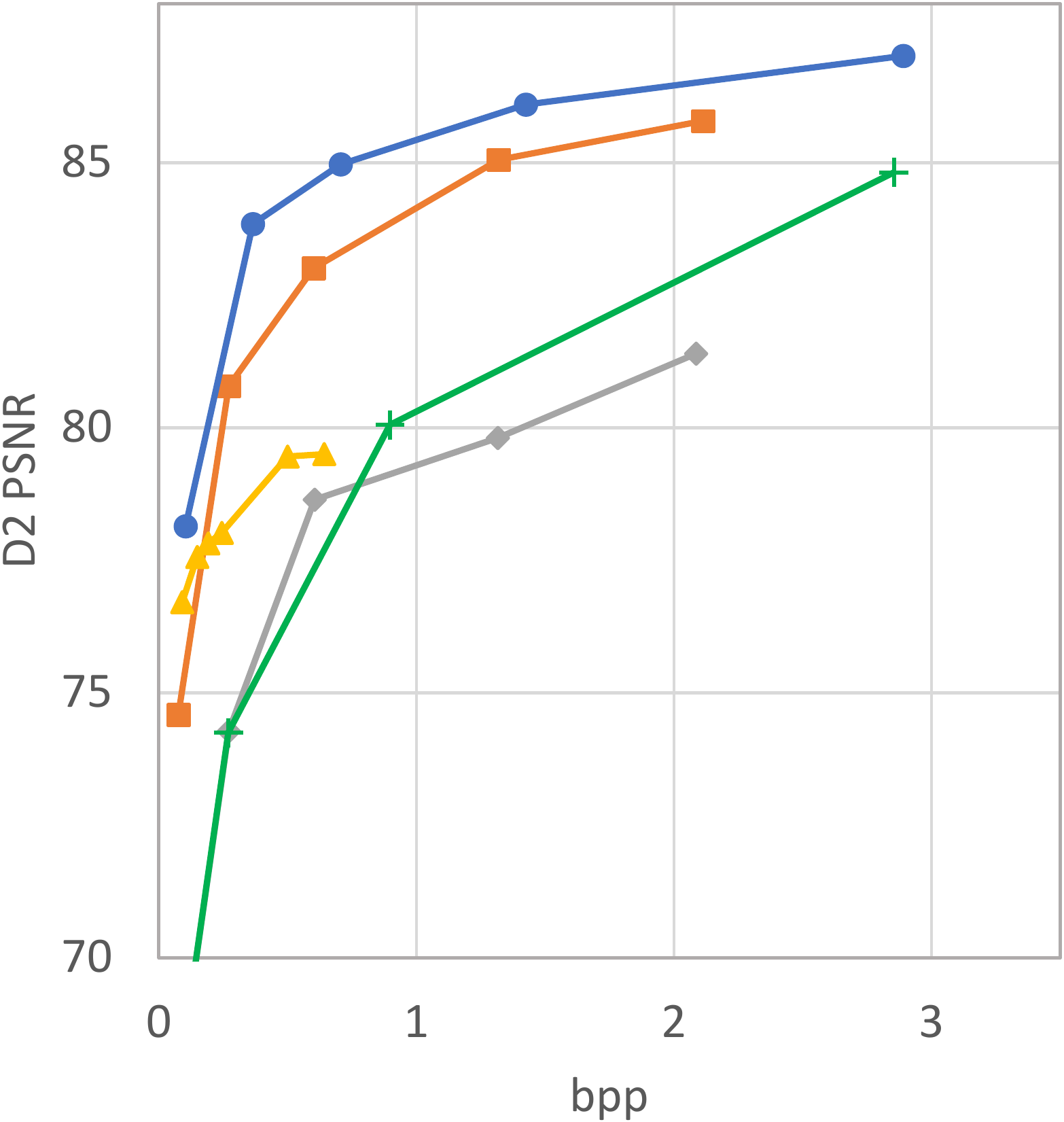}\label{fig:rd_boxer}}\hspace{5pt}
  \subfigure[``house\_without\_roof\_00057\_vox12'']{\includegraphics[height=120pt]{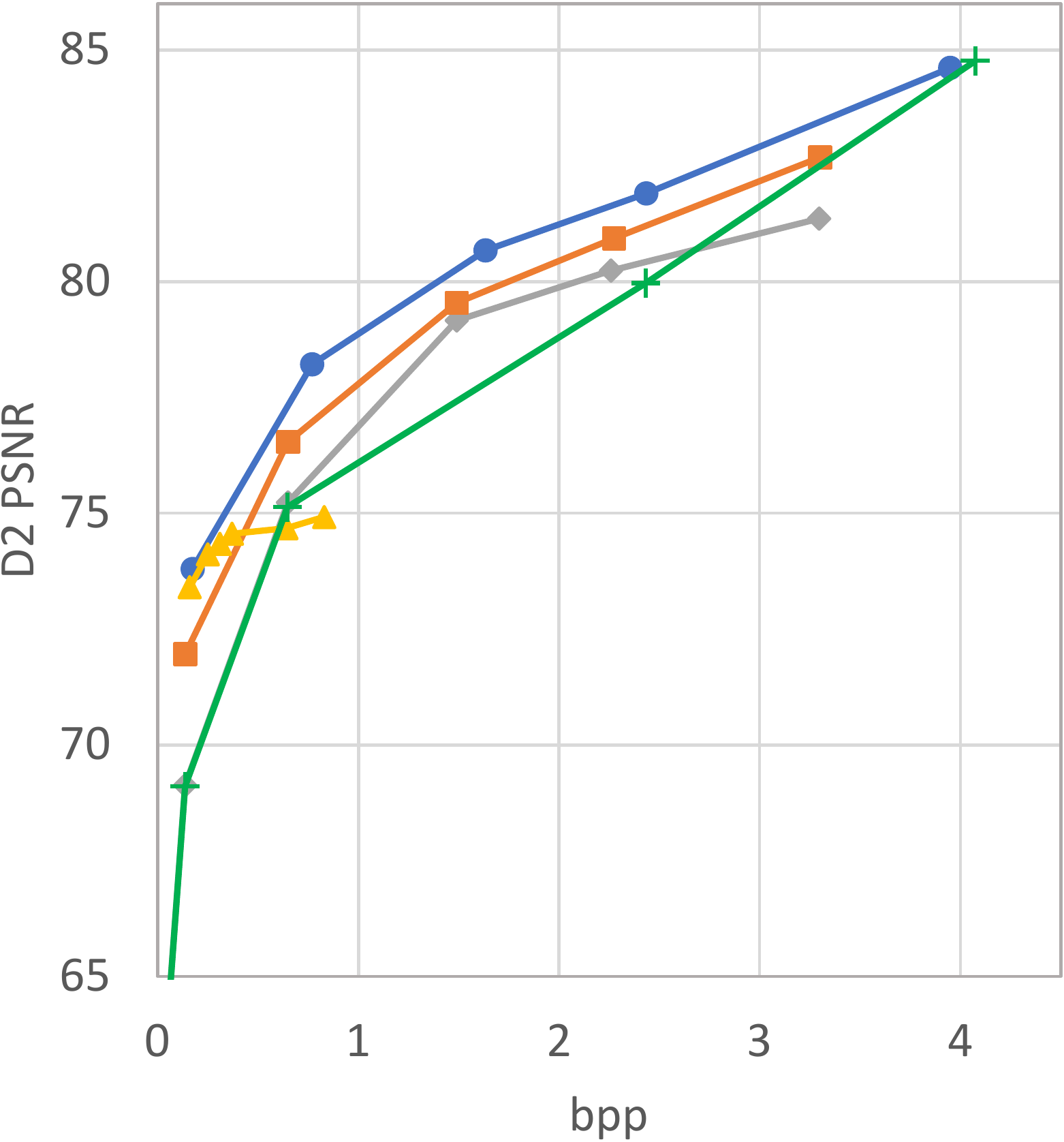}\label{fig:rd_house}}
  \caption{R-D performance of 12-bit dense point clouds, measured in both D1-PSNR and D2-PSNR. Our proposed GRASP-Net provides the best compression performance compared to the other methods.}
  \label{fig:rd_dense}
\end{figure*}

\begin{table*}[htbp]
  \centering\small
  \caption{BD-Rate and BD-PSNR gains against G-PCC octree (lossy) on the dense point clouds.}
    \begin{tabular}{c||cc|cc||cc|cc||cc|cc||cc|cc}
    \hline
    Method & \multicolumn{4}{c||}{PCGCv2}  & \multicolumn{4}{c||}{G-PCC+DUS} & \multicolumn{4}{c||}{GRASP-Net-Skip} & \multicolumn{4}{c}{GRASP-Net} \\
    \hline
    \multirow{2}[4]{*}{Metric} & \multicolumn{2}{c|}{BD-Rate} & \multicolumn{2}{c||}{BD-PSNR} & \multicolumn{2}{c|}{BD-Rate} & \multicolumn{2}{c||}{BD-PSNR} & \multicolumn{2}{c|}{BD-Rate} & \multicolumn{2}{c||}{BD-PSNR} & \multicolumn{2}{c|}{BD-Rate} & \multicolumn{2}{c}{BD-PSNR} \\
\cline{2-17}          & D1    & D2    & D1    & D2    & D1    & D2    & D1    & D2    & D1    & D2    & D1    & D2    & D1    & D2    & D1    & D2 \\
    \hline
    \hline
    Homo. & -54.26 & -62.69 & 2.72  & 3.80  & 6.61  & 8.18  & -0.20 & -0.32 & -42.64 & -62.96 & 2.31  & 4.01  & \textbf{-56.66} & \textbf{-85.46} & \textbf{3.90} & \textbf{6.11} \\
    Hetero. & -27.81 & -39.55 & 0.86  & 1.74  & -3.88 & -8.06 & 0.07  & 0.28  & -19.64 & -23.13 & 0.92  & 1.20  & \textbf{-34.60} & \textbf{-40.26} & \textbf{2.45} & \textbf{2.65} \\
    Avg.  & -44.34 & -54.01 & 2.02  & 3.02  & 2.68  & 2.09  & -0.10 & -0.09 & -34.01 & -48.02 & 1.79  & 2.95  & \textbf{-48.39} & \textbf{-68.51} & \textbf{3.35} & \textbf{4.81} \\
    \hline
    \end{tabular}%
  \label{tab:d1d2_dense}%
\end{table*}%

The decoded feature set $F_1'$ and the original one $F_1$ both contain pointwise features of $X_1$ describing local geometry, \ie, they both have $N$ features when $X_1$ has $N$ points.
However, the lengths of their features do not need to be the same, since $F_1'$ is not a decompressed version of $F_1$.
Our design can be extended to have fewer or more down-/up-sampling hierarchies under the same rationale, we empirically choose $2$ hierarchies (Fig.~\ref{fig:voxel_learning}) for dense point clouds while using $1$ hierarchy for sparse point clouds.

\section{Experimentation}
\label{sec:experiments}
This section demonstrates the state-of-the-art performance of GRASP-Net on both the sparse and the dense point clouds.

\subsection{Experimental Setup}
\textbf{Datasets}:
To verify the compression performance on dense point clouds, eight 12-bit point clouds are selected, including: ``longdress'', ``loot'', ``redandblack'', ``soldier'', ``house\_without\_roof'', ``boxer'', ``facade'', and ``frog''.
They are adopted by the MPEG Common Test Condition (CTC)~\cite{ctcgpcc} for geometry-based PCC, covering objects/scenes of various scales.
MPEG also classified ``house\_without\_roof'', ``facade'', and ``frog'' as (more challenging) heterogeneous point clouds; while the rest belong to the category of homogeneous point clouds \cite{mpeg2022dataset}.
To compress dense point clouds, our GRASP-Net is trained with the ModelNet40~\cite{wu20153d} dataset which contains 12k CAD models from $40$ categories of objects.

We employ the Ford dataset, which is also adopted by the MPEG CTC, for benchmarking on sparse point clouds.
The Ford dataset is collected by a spinning LiDAR mounted on a car driving in different scenes.
It contains $3$ dynamic point cloud sequences of 18-bit precision where each sequence contains $1500$ frames (or sweeps).
We use its first sequence ``Ford\_01\_q\_1mm'' with $1500$ point cloud frames for training and the rest $3000$ frames for benchmarking.

\textbf{Training}:
We end-to-end train our GRASP-Net by optimizing the rate-distortion (R-D) loss: $L = \alpha D + \beta R$, where $D$ measures the geometric distortion while $R$ is the bitrate of $B_\textrm{enh}$ estimated by the factorized prior model~\cite{balle2018variational}.
The parameters $\alpha$ and $\beta$ control the trade-off between distortion and bitrate.
To train a model for the skip mode, we simply let $L=D$.

\begin{figure*}[htbp]
  \centering \scriptsize 
  {
  \includegraphics[width=1.6\columnwidth]{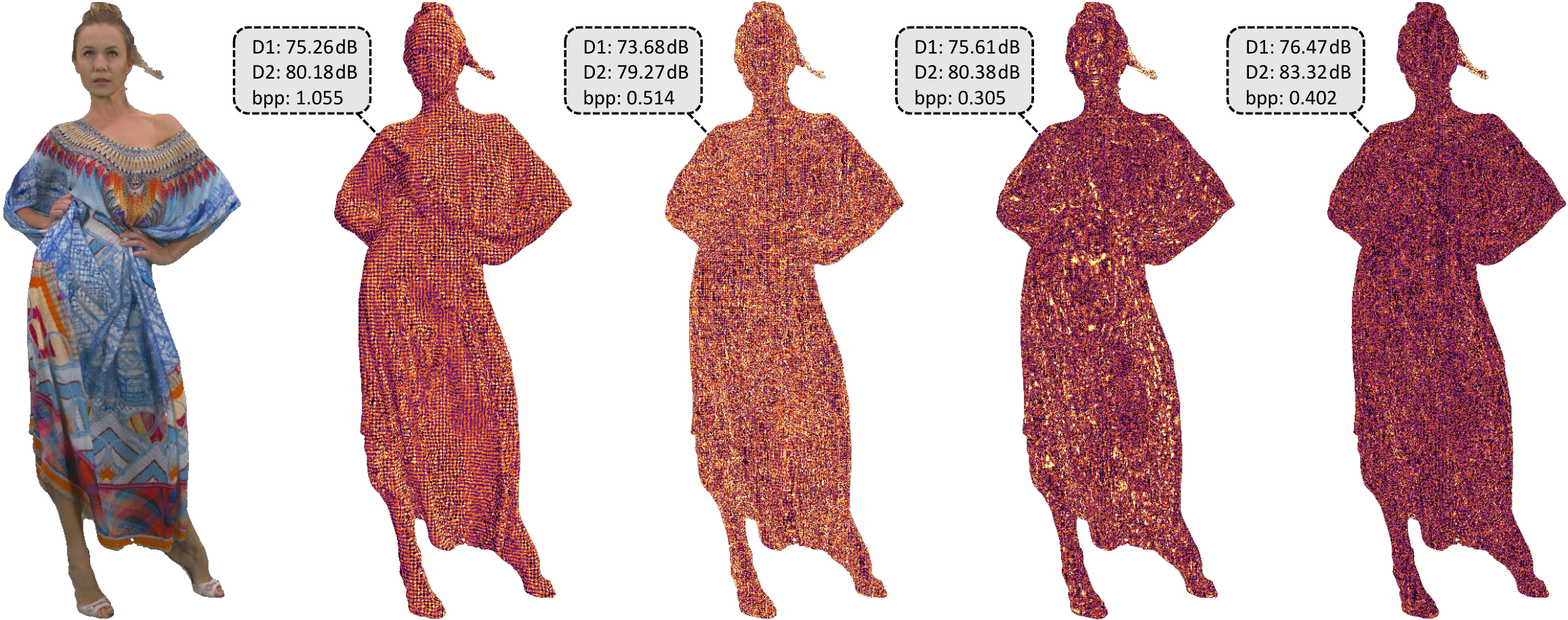}\hspace{15pt}
  \includegraphics[height=140pt]{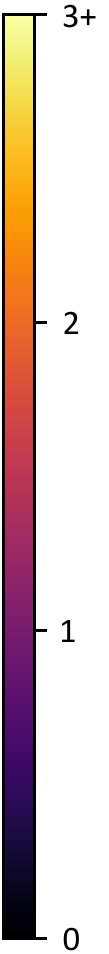}\\ \vspace{-5pt}
  \hspace{-25pt}
  \subfigure[Ground-truth]{\includegraphics[width=0.32\columnwidth]{figures/place_holder.png}}
  \subfigure[G-PCC]{\includegraphics[width=0.34\columnwidth]{figures/place_holder.png}\label{fig:dense_gpcc}}
  \subfigure[PCGCv2]{\includegraphics[width=0.35\columnwidth]{figures/place_holder.png}\label{fig:dense_pcgc}}
  \subfigure[GRASP-Net-Skip]{\includegraphics[width=0.35\columnwidth]{figures/place_holder.png}\label{fig:dense_skip}}
  \subfigure[GRASP-Net]{\includegraphics[width=0.32\columnwidth]{figures/place_holder.png}\label{fig:dense_grasp}}
  }
  \vspace{-10pt}
  \caption{Qualitative evaluation of ``longdress\_viewdep\_vox12''. The decoded point clouds are colored by the squared D1 errors.}
  \label{fig:vis_dense}
\end{figure*}

A decoded point cloud of our GRASP-Net is naturally a native 3D point set.
Thus, instead of computing $D$ with the cross entropy, \eg, \cite{wang2021multiscale,quach2020improved}, we use the \emph{augmented Chamfer distance}, as suggested in \cite{yang2018foldingnet, chen20203d}.
It is measured as:
\begin{equation*}\label{eq:chamfer}
  D(X, Y)=\max\left \{\frac{1}{|X|}\sum_{\mathbf{x} \in X}\min_{ \mathbf{y} \in  Y}  \left\| \mathbf{x} - \mathbf{y} \right\|_2,\frac{1}{|Y|}\sum_{ \mathbf{y} \in  Y} \min_{\mathbf{x} \in X} \left\| \mathbf{x} - \mathbf{y} \right\|_2\right \}
\end{equation*}
between two point clouds $X$ and $Y$, where $|\cdot|$ returns the number of points.

We use different quantization step sizes $s$ in the quantizer (Fig.~\ref{fig:framework}) to achieve different rate points.
For dense point clouds, we choose $s\in\{1.6,$ $2,$ $8/3,$ $4,$ $8\}$ for $5$ rate points; while for sparse point clouds, we choose $s\in\{2^6,$ $2^7,$ $2^8,$ $2^9,$ $2^{10}\}$ to have another $5$ rate points.
We fix $\alpha=5$, then for a different $s$, we empirically pick a value of $\beta$ for training to achieve the best R-D performance.
Intuitively, we look for a $\beta$ value that effectively improves the reconstruction quality while only introducing a small bitrate of $B_\textrm{enh}$.

\textbf{Implementation details}:
G-PCC octree (lossless)~\cite{graziosi2020overview} is employed as the octree coder of our base layer.
For the kNN search in the geometric subtraction module, we let $k=10$ when $s$ is $4$ or $8$ otherwise $k=5$.
We also remove those nearest points that are too far away from their query points.
On the decoder side, we let the Point Synthesis Network generates $k$ points for each feature vector, the same number of points as the kNN search.
Adam optimizer~\cite{kingma2014adam} is applied to train our GRASP-Net for $50$ epochs with the learning rate being $8\times10^{-4}$.
The batch size is $8$ when training on ModelNet40 while it is $2$ when training on Ford.
Our implementation is based on \emph{PccAI}~\cite{pccai}---an OSS infrastructure framework that facilitates the evaluation of learning-based PCC approaches.

\textbf{Benchmarking methods}:
We compare our GRASP-Net with state-of-the-art PCC methods: (i)~G-PCC octree (lossy)~\cite{graziosi2020overview} which is MPEG's standardized geometry-based PCC method; (ii)~PCGCv2~\cite{wang2021multiscale} dedicated for dense point cloud geometry compression with sparse CNN; and (iii)~VoxelContext-Net~\cite{que2021voxelcontext} accompanied with its coordinate refinement module, which performs well on sparse LiDAR point clouds.

Variants of GRASP-Net are also considered for ablation study: (i)~G-PCC+DUS (down-up scaling) is the base layer of GRASP-Net which uses MPEG G-PCC octree (lossless) as the octree coder; (ii)~GRASP-Net-Skip is the skip mode of GRASP-Net that operates without the enhancement bit-stream $B_\textrm{enh}$.

\begin{table}[htbp]
  \centering\small
  \caption{BD-Rate and BD-PSNR gains against G-PCC octree (lossy) on the Ford dataset, measured in both D1 and D2.}
    \begin{tabular}{c|c||c|c|c|c}
    \hline
    \multicolumn{2}{c||}{Metric} & {\scriptsize VoxelContext-Net} & {\scriptsize G-PCC+DUS} & {\scriptsize GRASP-Net-Skip} & {\scriptsize GRASP-Net} \\
    \hline
    \hline
    \multirow{2}[1]{*}{\footnotesize BD-Rate} & {\footnotesize D1}    & -10.32 & -2.08 & -19.53 & \textbf{-26.96} \\
          & D2    & -19.44 & -2.12 & -18.78 & \textbf{-34.87} \\
    \hline
    \multirow{2}[1]{*}{\footnotesize BD-PSNR} & {\footnotesize D1}    & 0.93  & 0.17  & 1.70  & \textbf{2.47} \\
          & D2    & 1.84  & 0.18  & 1.69  & \textbf{3.38} \\
    \hline
    \end{tabular}%
  \label{tab:d1d2_sparse}%
\end{table}%

Following the convention, we measure bitrate by \emph{bits per input point} (bpp); while distortion is measured by the peak signal-to-noise ratio (PSNR) based on \emph{point-to-point distance} (D1) and \emph{point-to-plane distance} (D2)~\cite{tian2017geometric}.
We also evaluate both BD-Rate and BD-PSNR gains by comparing two R-D curves, where the gains over MPEG G-PCC (lossy) are particularly interested.

\subsection{Performance Evaluation}
\textbf{Dense point clouds}:
Fig.~\ref{fig:rd_dense} plots several R-D curves achieved by our GRASP-Net on dense point clouds, along with those of the competing methods.
We also compute the BD-Rate and BD-PSNR gains using G-PCC as the baseline method, in which we ignore the smallest rate points to better quantify the overall performance across bitrates.
The results are presented in Table~\ref{tab:d1d2_dense}.

We see that our GRASP-Net provides the best R-D trade-off compared to other approaches.
PCGCv2~\cite{wang2021multiscale} performs well at small bitrate regime, though it fails as bitrate increases (Fig.~\ref{fig:rd_dense}).
That is because PCGCv2 is entirely based on convolutional layers.
It becomes less efficient for capturing the sparse local details that are critical at higher bitrates.
We also see that compared to GRASP-Net-Skip (skip mode), GRASP-Net achieves more gain.
It validates the effectiveness of introducing the enhancement bit-stream $B_\textrm{enh}$ dedicated for the geometric residual.

\begin{figure*}
  \centering \scriptsize 
  \hspace{0.3\columnwidth}\hspace{8pt}
  \subfigure{\includegraphics[width=0.38\columnwidth]{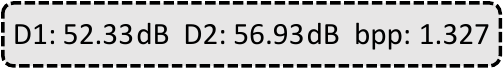}}\hspace{2pt}
  \subfigure{\includegraphics[width=0.38\columnwidth]{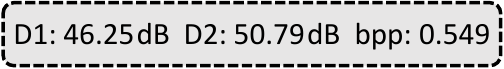}}\hspace{2pt}
  \subfigure{\includegraphics[width=0.38\columnwidth]{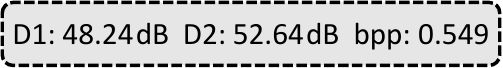}}\hspace{2pt}
  \subfigure{\includegraphics[width=0.38\columnwidth]{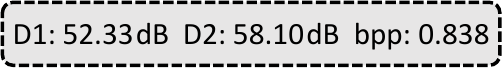}}\\\vspace{-10pt}
  \setcounter{subfigure}{0}
  \subfigure[Ground-truth]{\includegraphics[width=0.38\columnwidth]{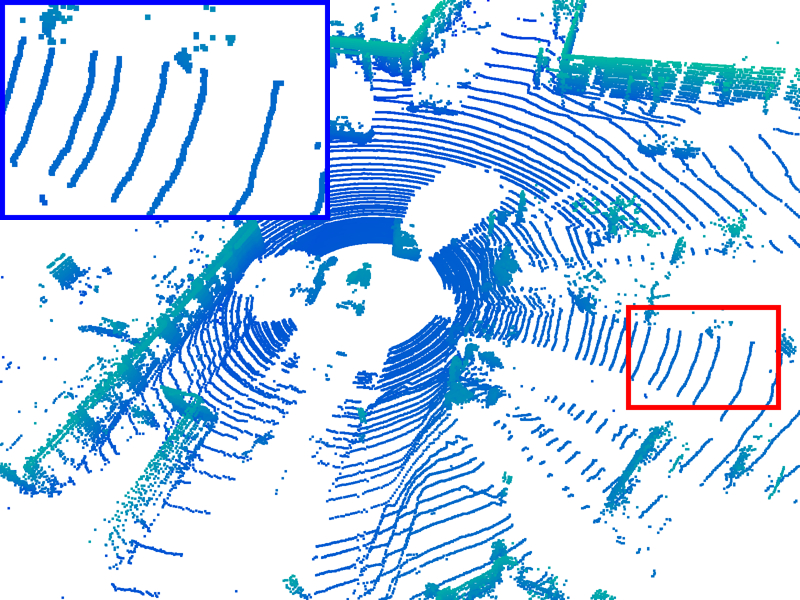}}\hspace{4pt}
  \subfigure[G-PCC]{\includegraphics[width=0.38\columnwidth]{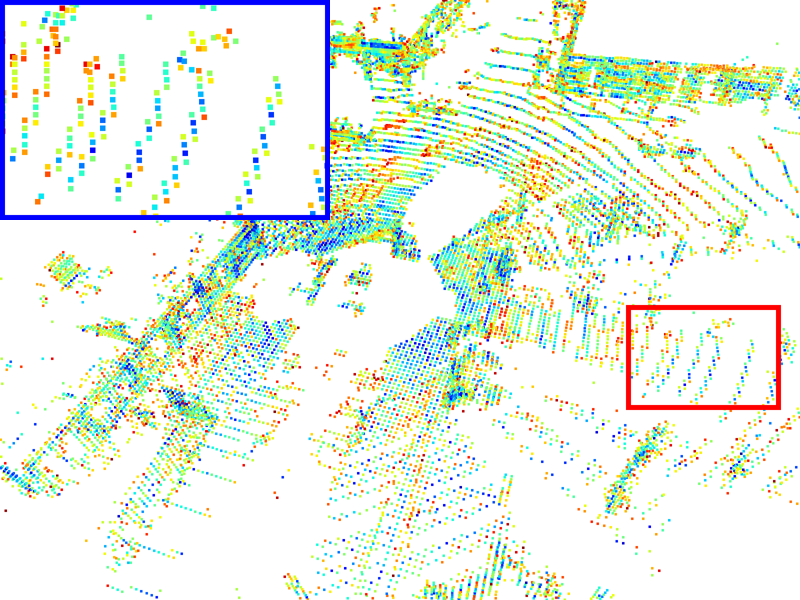}\label{fig:sparse_gpcc}}\hspace{2pt}
  \subfigure[G-PCC+DUS]{\includegraphics[width=0.38\columnwidth]{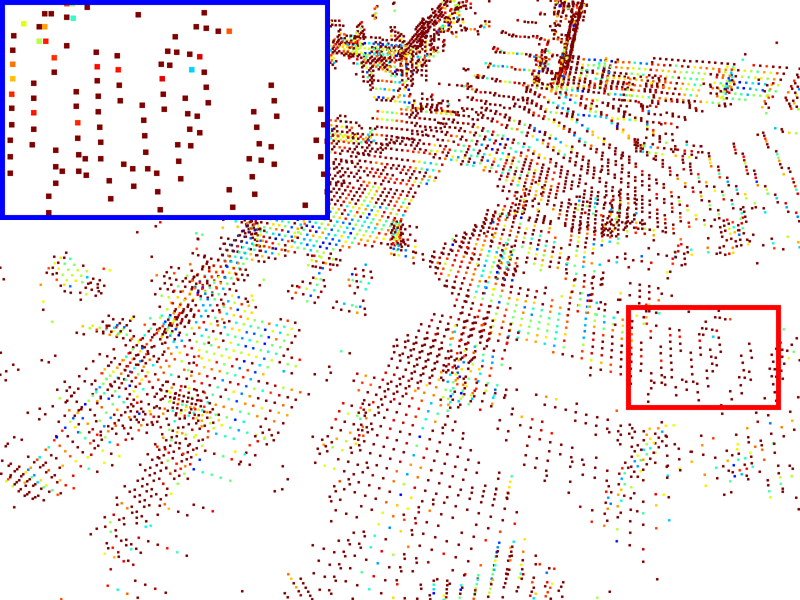}\label{fig:sparse_base}}\hspace{2pt}
  \subfigure[GRASP-Net-Skip]{\includegraphics[width=0.38\columnwidth]{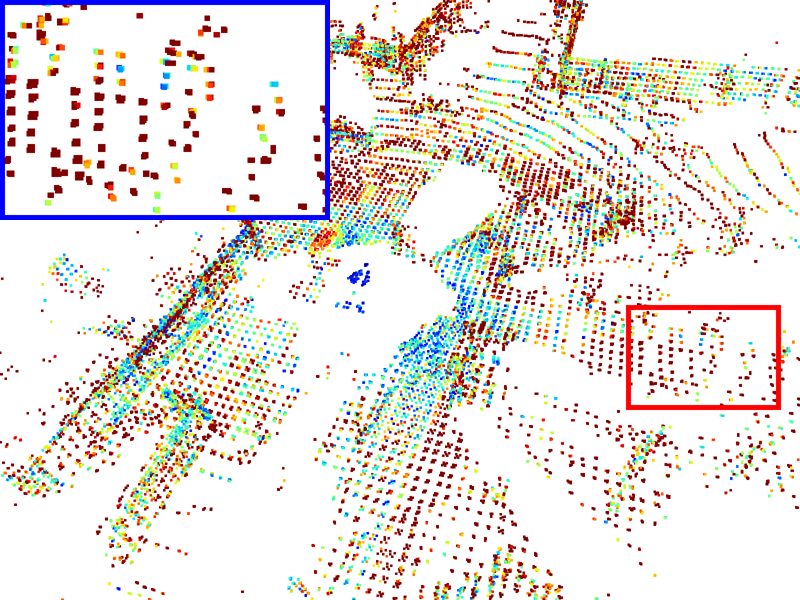}\label{fig:sparse_skip}}\hspace{2pt}
  \subfigure[GRASP-Net]{\includegraphics[width=0.38\columnwidth]{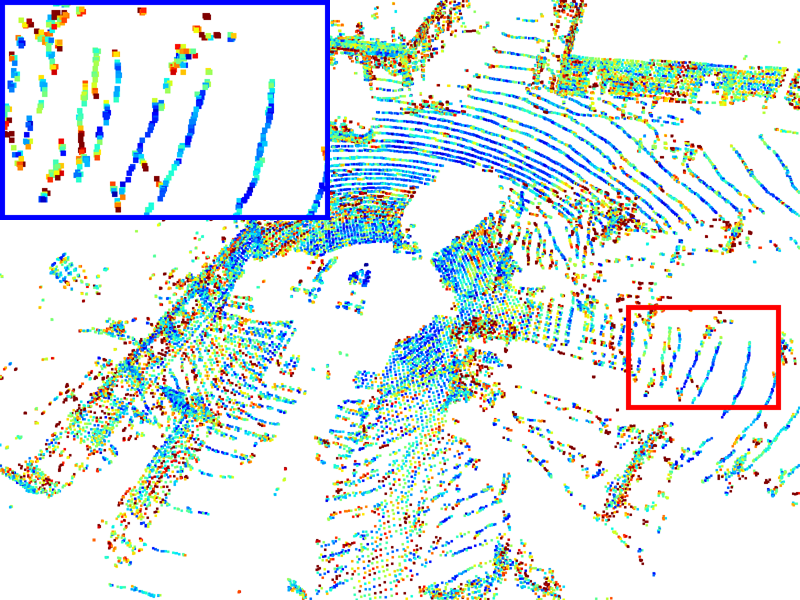}\label{fig:sparse_grasp}}
  \hspace{4pt}\includegraphics[height=70pt]{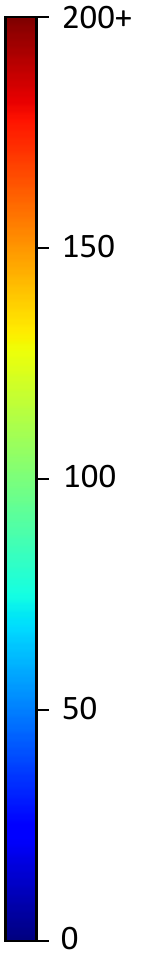}\\
  \vspace{-5pt}
  \caption{Qualitative evaluation of the point cloud frame ``Ford\_02\_vox1mm-0100''. The decoded point clouds are colored by the D1 error. The details in red boxes are zoomed in and shown in the blue boxes.}
  \label{fig:vis_sparse}
\end{figure*}

\begin{figure}
  \centering \scriptsize 
  \includegraphics[width=0.9\columnwidth]{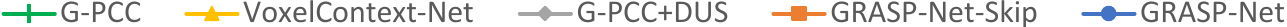}\\ \vspace{-5pt}
  \subfigure{\includegraphics[height=105pt]{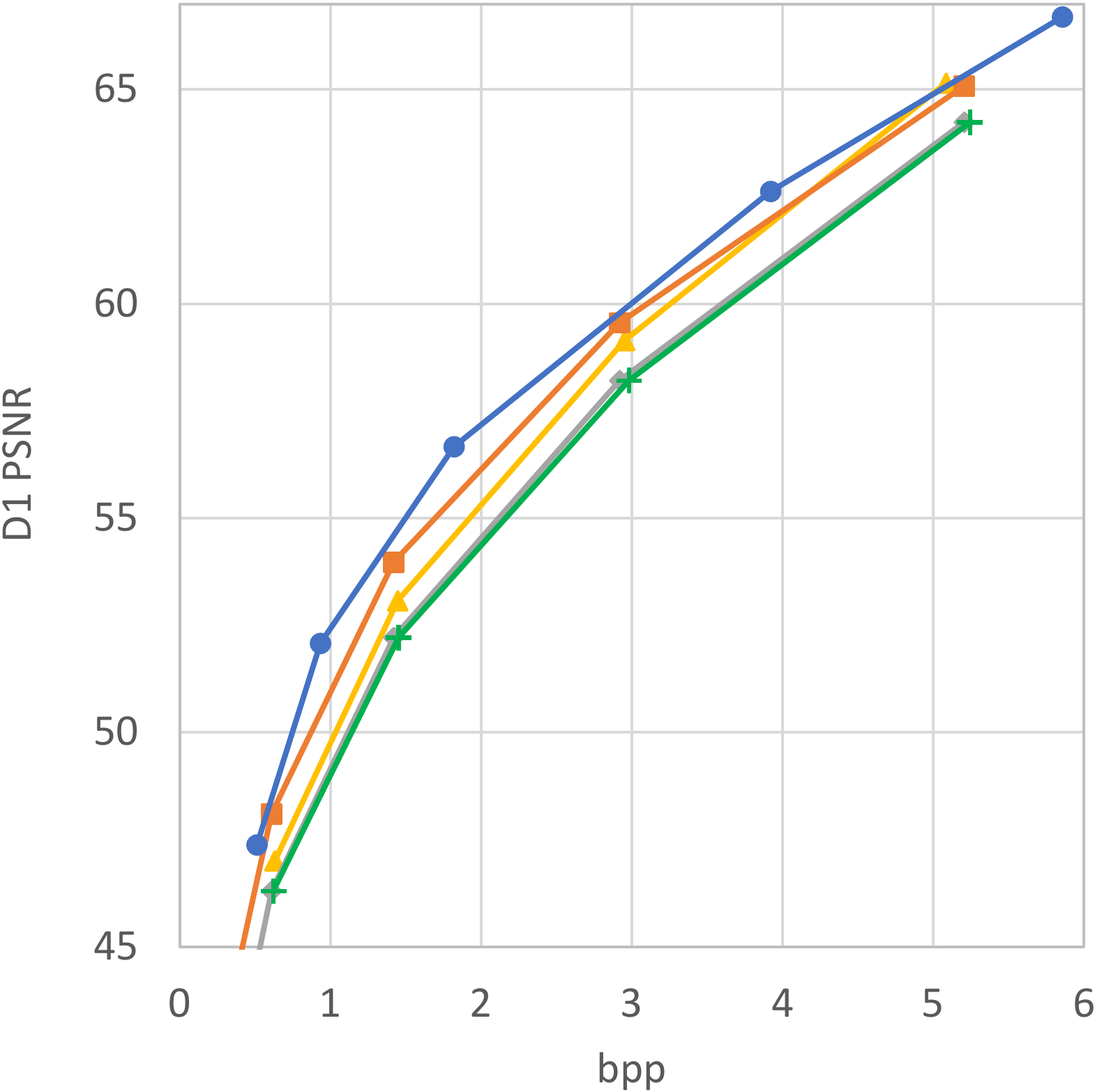}\label{fig:ford_d1}}\hspace{10pt}
  \subfigure{\includegraphics[height=105pt]{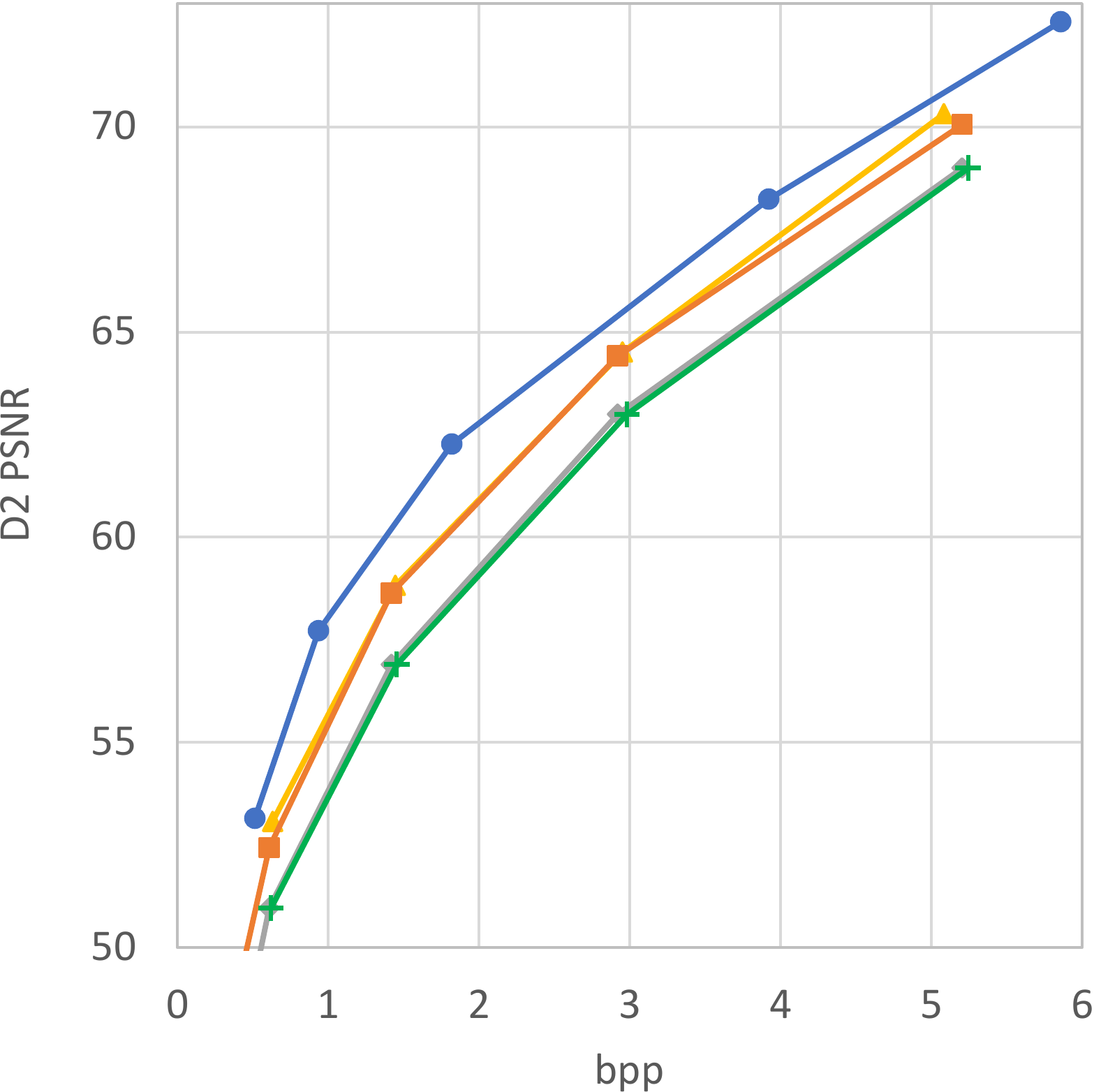}\label{fig:ford_d2}}
  \caption{R-D performance on the Ford sequences, measured in both D1-PSNR and D2-PSNR.}
  \vspace{-5pt}
  \label{fig:rd_sparse}
\end{figure}

From Table~\ref{tab:d1d2_dense} we see that for homogeneous point clouds, GRASP-Net brings about $30\%$ more gain in terms of D2 when compared to D1.
That is because the D2 metric of GRASP-Net ramps up much faster than D1 (Fig.~\ref{fig:rd_dense}a, b and c).
At around $1$~bpp, it reaches $85$~dB, comparable to the highest rate point ($>3$~bpp) of G-PCC.
It implies that at lower bitrates, the priority of GRASP-Net is to capture the object's surface.
Then based on the accurate knowledge of the surface, it devotes itself to improving the pointwise accuracy at higher bitrates.
On the other hand, the heterogeneous point clouds are more challenging due to their more erratic distributions.
Even for those cases, GRASP-Net still achieves a BD-Rate saving of about $-35\%$ in terms of D1.
On average, we bring a BD-Rate gain of about $-50\%$ on the 12-bit dense point clouds (Table~\ref{tab:d1d2_dense}).
Though Table~\ref{tab:d1d2_dense} shows PCGCv2 also performs well, it cannot reflect the failure of PCGCv2 at higher bitrates.
That is because the BD metrics only consider the overlapped part of the two compared curves in their computations~\cite{bdmetric}.

Qualitative comparisons are provided in Fig.~\ref{fig:vis_dense} where we visualize the decoded ``longdress'' point clouds of different methods by coloring them with the D1 squared error.
First, our result (Fig.~\ref{fig:dense_grasp}) has the best reconstruction quality at a much lower bitrate when compared to G-PCC and PCGCv2.
Moreover, compared to the skip mode (Fig.~\ref{fig:dense_skip}), the local details are greatly improved by introducing the enhancement bit-stream in GRASP-Net (Fig.~\ref{fig:dense_grasp}).

\textbf{Sparse point clouds}:
The R-D performance of our GRASP-Net and the competing methods on the Ford sequences are presented in Fig.~\ref{fig:rd_sparse}.
We see that GRASP-Net provides an obvious improvement over G-PCC in terms of both the D1 and D2 metrics.
It also out-performs the recently proposed VoxelContext-Net~\cite{que2021voxelcontext}.

The BD-Rate and BD-PSNR gains of different methods against G-PCC are calculated and presented in Table~\ref{tab:d1d2_sparse}, which clearly shows the advantage of the proposed GRASP-Net.
Without the enhancement bit-stream $B_\textrm{enh}$, the skip mode GRASP-Net-Skip already surpasses VoxelContext-Net in terms of D1.
Then by introducing $B_\textrm{enh}$, the proposed GRASP-Net further improves the performance, leading to an average BD-Rate gain of about $-27\%$ over G-PCC.

Fig.~\ref{fig:vis_sparse} visualizes the decoded point clouds of different methods by coloring them with the D1 error. 
We achieve a better reconstruction quality than G-PCC at a much lower bitrate.
Note how the reconstruction quality is improved from the base layer (Fig.~\ref{fig:sparse_base}) to the enhancement layer with skip mode (Fig.~\ref{fig:sparse_skip}), and to the case where $B_\textrm{enh}$ is used (Fig.~\ref{fig:sparse_grasp}).
And by inspecting Fig.~\ref{fig:vis_sparse} closely (\eg, the close-ups in the blue boxes), we see that GRASP-Net well reconstructs the intricate line patterns of the ground-truth.
It further confirms the benefits of GRASP-Net in preserving geometric details.

\subsection{Discussions}
\textbf{Complexity}:
Our GRASP-Net operates at a reasonable computational cost.
With a workstation with an Intel Xeon Gold 6234 CPU and an NVIDIA Quadro RTX 4000 GPU, our prototype on average takes $36.3$\,sec for encoding and $4.6$\,sec for decoding a 12-bit dense point cloud; while the encoding and decoding time for G-PCC is $30.0$\,sec and $1.5$\,sec, respectively. Though our runtime is longer than G-PCC, it can be greatly improved by having a more optimized implementation to accelerate the nearest neighbor search.
The number of parameters of our model (with $2$ hierarchies) is about 482k, which is much fewer than both PCGCv2 (778k) and VoxelContext-Net (2.15M).

\textbf{Bit allocation}:
GRASP-Net generates two bit-streams, $B_\textrm{bse}$ by the octree coder, and $B_\textrm{enh}$ by the enhancement layer.
In all cases, $B_\textrm{bse}$ dominates and occupies at least $60\%$ of bits.
This reveals a potential improvement to GRASP-Net, that is to employ a more advanced octree coder in the base layer, \eg, OctSqueeze~\cite{huang2020octsqueeze} and SparsePCGC~\cite{wang2021sparse} codecs.
We left this for future investigation.

\section{Conclusion}
\label{sec:conclusion}
The local sparsity of real-world point clouds is analyzed by inspecting their dimensions across different scales.
Motivated by this phenomenon, we propose Geometric Residual Analysis and Synthesis for PCC (GRASP-Net)---a heterogeneous PCC framework based on deep learning.
Especially, to compress the challenging geometric details, we adopt native point-based learning to convert 3D points into more regularly-distributed latent features; then sparse convolutional layers operating on voxel grids are applied to capture the feature correlation.
The state-of-the-art performance of GRASP-Net is proven via extensive experiments on both dense and sparse point clouds.
A future research direction is to generalize our heterogeneous framework to address other point cloud processing tasks, \eg, denoising, segmentation, and classification.

\bibliographystyle{ACM-Reference-Format}
\balance



\end{document}